\documentclass[11pt,a4paper]{article}
\usepackage{jcappub}

\usepackage{textpos}
\usepackage{subcaption}
\usepackage{appendix}

\usepackage{hyperref}
\hypersetup{
    colorlinks=true,
    linkcolor=blue,
    citecolor=blue,
}
\usepackage{bm}

\def\be{\begin{equation}}
\def\ee{\end{equation}}
\begin{document}

%%%%%%%%%%%%%%%%%%%%%%%%%%%%%%%%%%%%%%%%%%%%%%%%%%%%%%%%%%%%%%%%%%%
% Title Page
%%%%%%%%%%%%%%%%%%%%%%%%%%%%%%%%%%%%%%%%%%%%%%%%%%%%%%%%%%%%%%%%%%%

\title{The variation of the fine-structure constant from disformal couplings}
\author[a]{Carsten van de Bruck}
\author[a]{,~Jurgen Mifsud}
\author[b]{and Nelson J. Nunes}
\affiliation[a]{Consortium for Fundamental Physics, School of Mathematics and Statistics, University of Sheffield, Hounsfield Road, Sheffield S3 7RH, UK}
\affiliation[b]{Instituto de Astrof\'isica e Ci\^encias do Espa\c{c}o, Faculdade de Ci\^encias da Universidade de Lisboa, Campo Grande, PT1749-016 Lisboa, Portugal} 
\emailAdd{c.vandebruck@sheffield.ac.uk}
\emailAdd{jmifsud1@sheffield.ac.uk}
\emailAdd{njnunes@fc.ul.pt}

\abstract{We study a theory in which the electromagnetic field is disformally coupled to a scalar field, in addition to a usual non--minimal electromagnetic coupling. We show that disformal couplings modify the expression for the fine--structure constant, $\alpha$. As a result, the theory we consider can explain the non--zero reported variation in the evolution of $\alpha$ by purely considering disformal couplings. We also find that if matter and photons are coupled in the same way to the scalar field, disformal couplings itself do not lead to a variation of the fine--structure constant. A number of scenarios are discussed consistent with the current astrophysical, geochemical, laboratory and the cosmic microwave background radiation constraints on the cosmological evolution of $\alpha$. The models presented are also consistent with the current type Ia supernovae constraints on the effective dark energy equation of state. 
We find that the Oklo bound in particular puts strong constraints on the model parameters. From our numerical results, we find that the introduction of a non--minimal electromagnetic coupling enhances the cosmological variation in $\alpha$. Better constrained data is expected to be reported by ALMA and with the forthcoming generation of high--resolution ultra--stable spectrographs such as PEPSI, ESPRESSO, and ELT--HIRES. Furthermore, an expected increase in the sensitivity of molecular and nuclear 
clocks will put a more stringent constraint on the theory.}

\maketitle

%%%%%%%%%%%%
\section{Introduction}
%%%%%%%%%%%%
\hspace{1cm}The idea on the variation of the fundamental constants of physics was first raised by Dirac in his `large numbers hypothesis' \cite{Dirac}. The advent of higher-dimensional theories \cite{Chodos,Wu}, in which the effective (3+1)-dimensional constants can vary in space and time, has led to {an increased} interest in nature's fundamental constants and their variation. Despite this, Einstein's equivalence principle, and hence local position invariance, is one of the building blocks of gravitational metric theories, which include amongst others the well-known gravitational theory of general relativity. Hence, the link between the constancy of fundamental constants and the equivalence principle is a natural one. 
\par
Recent observations have shown the possibility of the variation of one of these fundamental constants of nature, the electromagnetic fine-structure constant, $\alpha=e^2/4\pi\epsilon_0\hbar c\simeq1/137$ \cite{CODATA}. Weaker constraints on the variation of the electromagnetic fine-structure constant have been derived from Big Bang Nucleosynthesis (BBN) at $z\sim10^9$ \cite{Iocco} and the cosmic microwave background (CMB) radiation at $z\sim10^3$ \cite{P1}. More stringent bounds have been reported at lower redshifts from astrophysical \cite{E,F,G,H,I,J,K,N,King,Rahmani,Murphy04,AD}, geochemical \cite{O,Fujii,Davis} and laboratory \cite{M} constraints.   
\par
In this work, we will be considering a theoretical model containing three distinct metrics, related by a disformal transformation. We will specify our scalar-gravitational sector in what we will later call the Einstein Frame, in which the action looks like general relativity with an added scalar field. We then specify two other distinct frames in which we define matter and radiation separately. The latter two frames are described by two separate metrics, $\tilde{g}^{(i)}_{\mu\nu}$, although both of these metrics are disformally related with the scalar-gravitational metric, $g_{\mu\nu}$. Such disformal transformations have been introduced by Bekenstein \cite{Bekenstein}, where in general a disformal transformation is characterized by the following relation
\begin{equation}\label{disformal_relation}
\tilde{g}^{(i)}_{\mu\nu}=C_i(\phi,X)g_{\mu\nu}+D_i(\phi,X)\phi_{,\mu}\phi_{,\nu}
\end{equation}
where $X=(1/2)g^{\mu\nu}\phi_{,\mu}\phi_{,\nu}$ is the kinetic term. The first term is the widely known conformal coupling, whereas the second term is referred to as the disformal coupling. In what follows, we will be considering both couplings to depend on $\phi$ only. The line elements of these disformal metrics are now related by $d\tilde{s}_i^2=C_ids^2+D_i(\phi_{,\mu}dx^\mu)^2$, and as a result null four-vectors with respect to $g_{\mu\nu}$ could be either spacelike or timelike four-vectors with respect to the disformal metric $\tilde{g}^{(i)}_{\mu\nu}$. Furthermore, a conformal transformation can be viewed as an angle preserving metric rescaling mapping, although a disformal transformation deforms the spacetime to a preferred direction characterized by the gradient of the fields resulting in a distortion of both angles and lengths \cite{Koivisto}. 
\par
Disformal relations as described in (\ref{disformal_relation}) have been used in different scenarios, including inflation \cite{Kaloper}, varying speed of light theories \cite{Clayton,Magueijo,Bassett}, massive gravity \cite{Rham1,Rham2}, dark energy \cite{Koivisto1,Zuma5,Sakstein:2014aca,Sakstein:2014isa} and others. It was shown in \cite{Bettoni:2013diz} that disformal transformations preserve second order field equations and play, in the Horndeski theory, a similar role conformal transformations play in the scalar-tensor theories. See e.g. \cite{Brax:2014vva} on current constraints on universal disformal couplings.  
\par
The most immediate way to obtain a space and time variation of the electromagnetic fine-structure constant is to introduce a non--minimal coupling between the scalar field and electromagnetic fields \cite{MM, Bekenstein1,Barrow,Barrow1,Barrow2}. It has been shown that such a non--minimal coupling give rise to nonconservation of the photon number along geodesics \cite{Minazzoli2} which result in a modification of the distance-duality relation \cite{MM,Minazzoli1,Etherington1,Etherington2,Ellis1,Ellis2,Ellis3} and CMB spectrum distortion \cite{MM,Lima1,Lima2}. Similar results have also been reported in disformal scalar-tensor theories \cite{Carsten,Brax}. We will show that such non--minimal coupling is not necessary to explain this non-zero variation of $\alpha$, as we can reproduce a varying electromagnetic fine-structure coupling purely by the introduced disformally related metrics. 
\par
We introduce our model in Section \ref{sec:Model} and present the cosmological equations in Section \ref{sec:Cosmology} in which we give the relevant equations that we then use for our examples. In Section \ref{sec:constraints} we list the observational constraints that we use in order to choose our model parameters, which we then present and discuss in Section \ref{sec:analysis}.

%%%%%%%%%%%%
\section{Disformal Electrodynamics: The Model}\label{sec:Model}
%%%%%%%%%%%%
We consider the following action, consisting of a gravitational sector, a matter sector and an electromagnetic sector, respectively: 
\begin{eqnarray}
{\cal S} = {\cal S}_{\rm grav}\left(g_{\mu\nu},\phi \right) + S_{\rm matter} ({\tilde g}_{\mu\nu}^{(m)}) + S_{\rm EM} (A_\mu, {\tilde g}_{\mu\nu}^{(r)}).
\end{eqnarray}
The metrics ${\tilde g}_{\mu\nu}^{(m)}$ and ${\tilde g}_{\mu\nu}^{(r)}$ are related to $g_{\mu\nu}$ via a disformal transformation:
\begin{align}
{\tilde g}_{\mu\nu}^{(m)} &= C_{m} g_{\mu\nu} + D_m \phi_{,\mu}\phi_{,\nu} \label{g_m} ~, \\
{\tilde g}_{\mu\nu}^{(r)} &= C_{r} g_{\mu\nu} + D_r \phi_{,\mu}\phi_{,\nu} \label{g_r}~.
\end{align}
Here, $C_{r,m}$ and $D_{r,m}$ are functions of the scalar field. The functions $C_{r,m}$ are conformal factors, whereas the functions $D_{r,m}$ are disformal couplings. At this point, we do not specify the gravitational sector. Instead, we study the consequences of the disformally coupled scalar field on the propagation of electromagnetic waves. We aim to work in the Jordan Frame, by which we mean the frame in which matter is decoupled from the scalar degree of freedom. We therefore perform a disformal transformation of the action above such that all parts of the action are written in terms of the metric $\tilde{g}_{\mu\nu}^{(m)}$. The electromagnetic sector is specified by 
\begin{equation}
{\cal S}_{\rm EM} = -\frac{1}{4} \int d^4 x \sqrt{-\tilde{g}^{(r)}} h(\phi) \tilde{g}_{(r)}^{\mu\nu} \tilde{g}_{(r)}^{\alpha\beta} F_{\mu\alpha}F_{\nu\beta} -\int d^4 x \sqrt{-\tilde{g}^{(m)}} \tilde{g}_{(m)}^{\mu\nu}j_\mu A_\mu,
\end{equation}
where $F_{\mu\nu}$ is the standard antisymmetric Faraday tensor, $F_{\mu\nu}= \partial_\mu A_\nu - \partial_\nu A_\mu$, and $j^\mu$ is the four--current. The function $h(\phi)$ is the direct coupling between the electromagnetic field and the scalar. 
Note that $\tilde{g}^{(r)}_{\mu\nu}$ can also be written as
\begin{equation}\label{rad}
\tilde{g}^{(r)}_{\mu\nu} = \frac{C_r}{C_m}\tilde{g}_{\mu\nu}^{(m)} + \left( D_r - \frac{C_r D_m}{C_m}  \right)\phi_{,\mu}\phi_{,\nu} \equiv A \tilde{g}^{(m)}_{\mu\nu} + B\phi_{,\mu}\phi_{,\nu} ,
\end{equation}
Then, in terms of this metric the electromagnetic sector becomes
\begin{equation}\label{EMaction}
\begin{split}
{\cal S}_{\rm EM} =& -\frac{1}{4} \int d^4 x \sqrt{-\tilde{g}^{(m)}}h(\phi) Z\left[\tilde{g}^{\mu\nu}_{(m)}\tilde{g}^{\alpha\beta}_{(m)} - 2 \gamma^2 \tilde{g}^{\mu\nu}_{(m)}\phi^{,\alpha}\phi^{,\beta} \right]F_{\mu\alpha}F_{\nu\beta}~\\
& -\int d^4 x \sqrt{-\tilde{g}^{(m)}} {\tilde g}_{(m)}^{\mu\nu}j_\mu A_\mu~,
\end{split}
\end{equation}
where we raise the indices with the metric $\tilde{g}^{(m)}_{\mu\nu}$ and define
\begin{equation}\label{Zdefinition}
Z = \left( 1 + \frac{B}{A}\tilde{g}^{\mu\nu}_{(m)}\partial_\mu \phi \partial_\nu \phi  \right)^{1/2},
\end{equation}
together with
\begin{equation}
\gamma^2 = \frac{B}{A+B\tilde{g}^{\mu\nu}_{(m)}\partial_\mu \phi \partial_\nu \phi}~.
\end{equation}
Note that the term proportional to $\gamma^4$ vanishes due to the antisymmetry of $F_{\mu\nu}$. Furthermore, note that gauge invariance implies $\tilde{\nabla}_\mu j^\mu = 0$, where the covariant derivative is compatible with the metric $\tilde{g}_{\mu\nu}^{(m)}$. 
The field equations can be readily obtained by varying the action with respect to $A_\mu$,
which results in
\begin{eqnarray}\label{generalfield}
\tilde{\nabla}_\epsilon\left(h(\phi) Z F^{\epsilon\rho} \right) - \tilde{\nabla}_\epsilon \left( h(\phi) Z\gamma^2 \phi^{,\beta}\left( \tilde{g}_{(m)}^{\epsilon\nu}\phi^{,\rho} -  \tilde{g}_{(m)}^{\rho\nu}\phi^{,\epsilon} \right)F_{\nu\beta} \right) = j^\rho~.
\end{eqnarray}
where we again raise the indices with $\tilde{g}_{\mu\nu}^{(m)}$. The first term in the action (\ref{EMaction}) contains two parts: the first part consists of the kinetic term for the vector potential $A_\mu$; the second part is an interaction term between the disformally coupled scalar field and $A_\mu$. The latter vanishes in the case of vanishing disformal couplings. 
From the form of the action (\ref{EMaction}) we might naively think that the fine-structure coupling is simply given by $\alpha \propto 1/hZ$. This is not the case, as we shall show. 
To identify the effective electromagnetic coupling (or the {"fine--structure constant"}) $\alpha$, we start by deriving the field equation for the electric field in Minkowski space $\left(\tilde{g}_{\mu\nu}^{(m)}=\eta_{\mu\nu}\right)$ and set the bare speed of light $c=1$. We also consider the scalar field to be a function of time only. From the obtained field equation (\ref{generalfield}), and using the fact that the electric field is identified by, $E^i=F^{i0}$, we find that the field equation for the electric field is given by 
\begin{equation}
\nabla \cdot {\bf E} = \frac{Z\rho}{h(\phi)}
\end{equation}
where $\rho= j^0$ is the charge density. By integrating this equation over a volume ${\cal V}$, it is straightforward to derive the electrostatic potential $V$ (for which ${\bf E} = -\nabla V$), which is found to be $V(r) = ZQ/(4\pi h(\phi) r)$, where $Q$ is the total charge contained in ${\cal V}$. Comparing this to the standard expression for the tree-level-potential from QED, one finds that $\alpha$ has the following dependence on $Z$ and $h$: 
\begin{equation}
\alpha \propto \frac{Z}{h(\phi)}. 
\end{equation}
Note that if matter and radiation couple in the same way to the scalar field (i.e. ${\tilde g}_{\mu\nu}^{(m)}= {\tilde g}_{\mu\nu}^{(r)}$ for which $A = 1$ and $B=0$) we have $Z=1$ and we recover the usual form $\alpha\propto h(\phi)^{-1}$ \cite{MM,Dam1,Dam2,Dam3,Dam4,Nunes1,Olive,Nunes2,Uzan}.  

%%%%%%%%%%%%
\subsection{Cosmology}\label{sec:Cosmology}
%%%%%%%%%%%%
We now specify our gravitational-scalar action to be as follows
\begin{eqnarray}
{\cal S}_{\rm grav}\left(g_{\mu\nu},\phi \right) = \int d^4x\sqrt{-g}\left(\frac{M_{\rm Pl}^2}{2}R-\frac{1}{2}g^{\mu\nu}\partial_{\mu}\phi\partial_{\nu}\phi-V(\phi)\right)
\end{eqnarray}
where $R$ is the Ricci scalar calculated with respect to the metric $g_{\mu\nu}$. From now on, reduced Planck units are assumed: $M_{\rm Pl}=1$. Hence, the theory we consider is given in the Einstein Frame as
\begin{equation}\label{fullaction}
\begin{split}
{\cal S} =& \int d^4x\sqrt{-g}\left(\frac{1}{2}R-\frac{1}{2}g^{\mu\nu}\partial_{\mu}\phi\partial_{\nu}\phi-V(\phi)\right) +  S_{\rm matter} ({\tilde g}_{\mu\nu}^{(m)})\\
& -\frac{1}{4} \int d^4 x \sqrt{-\tilde{g}^{(r)}} h(\phi) \tilde{g}_{(r)}^{\mu\nu} \tilde{g}_{(r)}^{\alpha\beta} F_{\mu\alpha}F_{\nu\beta}.
\end{split}
\end{equation}
The last term in the action above describes the dynamics of the CMB photons. 
From now on, we will refer to the three distinctive frames as the Einstein Frame (EF), Radiation Frame (RF) and Jordan Frame (JF) corresponding to the metrics $g_{\mu\nu},\;\tilde{g}_{\mu\nu}^{(r)}$ and $\tilde{g}_{\mu\nu}^{(m)}$ respectively. We define the RF as the frame in which all electromagnetic quantities are defined in their standard way. Furthermore, we define the JF as the frame in which matter is uncoupled from the scalar field. Since the gravity-scalar part of the action is written in its simplest form in the EF, we will be working in this frame and not in the other two frames where in general this part of the action has a nonstandard form. 
\par
By the variation of the action (\ref{fullaction}) with respect to the metric $g_{\mu\nu}$, we obtain the Einstein Field Equations
\begin{equation}\label{EFE}
G^{\mu\nu}=T^{\mu\nu}_{\phi}+T^{\mu\nu}_{(m)}+T^{\mu\nu}_{(r)} ,
\end{equation}
with $G^{\mu\nu}=R^{\mu\nu}-\frac{1}{2}g^{\mu\nu}R$ being the usual Einstein tensor in the EF. The energy-momentum tensors of the scalar field, matter and radiation are denoted by $T^{\mu\nu}_{\phi},\;T^{\mu\nu}_{(m)}$ and $T^{\mu\nu}_{(r)}$ respectively. We specify these  energy-momentum tensors to be as follows
\begin{align}
T_{\mu\nu}^{\phi}&=\partial_\mu\phi\partial_\nu\phi-g_{\mu\nu}\left(\frac{1}{2}g^{\rho\sigma}\partial_\rho\phi\partial_\sigma\phi+V(\phi)\right)\label{T_phi} , \\
T^{(m)}_{\mu\nu}&=-\frac{2}{\sqrt{-g}}\frac{\delta(\sqrt{-\tilde{g}^{(m)}}\tilde{\mathcal{L}}_m)}{\delta g^{\mu\nu}}, \\
T^{(r)}_{\mu\nu}&=-\frac{2}{\sqrt{-g}}\frac{\delta(\sqrt{-\tilde{g}^{(r)}}\tilde{\mathcal{L}}_{EM})}{\delta g^{\mu\nu}},
\end{align}
where we define the electromagnetic Lagrangian by $\tilde{\mathcal{L}}_{EM}=-\frac{1}{4}h(\phi)\tilde{g}_{(r)}^{\mu\nu} \tilde{g}_{(r)}^{\alpha\beta} F_{\mu\alpha}F_{\nu\beta}$ and denote the matter Lagrangian by $\mathcal{\tilde{L}}_m$. We now use the variation of the action (\ref{fullaction}) with respect to the scalar field, which leads us to the Klein-Gordon equation
\begin{equation}\label{klein-gordon}
\square\phi-V^{\prime}=-Q_m-Q_r ,
\end{equation}
where we have introduced a matter coupling strength, $Q_m$, and a radiation coupling strength, $Q_r$, as follows
\begin{align}
Q_m&=\frac{C_m^{\prime}}{2C_m}T_{(m)}+\frac{D_m^{\prime}}{2C_m}\phi_{,\mu}\phi_{,\nu}T^{\mu\nu}_{(m)}-\nabla_\mu\left[\frac{D_m}{C_m}\phi_{,\nu}T^{\mu\nu}_{(m)}\right]\label{Qmnew} ,\\
Q_r&=\frac{C_r^{\prime}}{2C_r}T_{(r)}+\frac{D_r^{\prime}}{2C_r}\phi_{,\mu}\phi_{,\nu}T^{\mu\nu}_{(r)}+\frac{h^{\prime}}{h}C_r^2\sqrt{1+\frac{D_r}{C_r}g^{\mu\nu}\phi_{,\mu}\phi_{,\nu}}\;\mathcal{\tilde{L}}_{EM}-\nabla_\mu\left[\frac{D_r}{C_r}\phi_{,\nu}T^{\mu\nu}_{(r)}\right]\label{Q_r_def}, 
\end{align}
where $T_{(r)}$ and $T_{(m)}$ are the trace of $T^{\mu\nu}_{(r)}$ and $T^{\mu\nu}_{(m)}$ respectively, and in (\ref{Q_r_def}) we have also used (\ref{det_g}). As a result of the Bianchi identities, the total energy-momentum tensor in the EF is covariantly conserved with respect to the EF metric, leading to the following conservation relation
\begin{equation}\label{relation}
\nabla_\mu(T^{\mu\nu}_\phi+T^{\mu\nu}_{(m)}+T^{\mu\nu}_{(r)})=0.
\end{equation}
Although this holds for the total energy-momentum tensor, the couplings under consideration 
do not allow each of the energy-momentum tensors to be individually conserved.
 Indeed, by using (\ref{T_phi}), (\ref{klein-gordon}) and (\ref{relation}), we find that the matter and radiation conservation equations now read as follows 
\begin{align}
\nabla_\mu T^{\mu}_{(m)\nu}&=Q_m\phi_{,\nu}\label{matter_cons}, \\
\nabla_\mu T^{\mu}_{(r)\nu}&=Q_r\phi_{,\nu}\label{radiation_cons}.
\end{align}
A similar derivation is described in \cite{Carsten,Zuma1,Zuma2,Minamitsuji,Zuma3}. We shall now consider perfect fluid energy-momentum tensors for radiation and matter in the EF 
\begin{align}
T^{\mu\nu}_{(r)}&=(\rho_r+p_r)u^{\mu}u^{\nu}+p_rg^{\mu\nu}\label{TEF} , \\
T^{\mu\nu}_{(m)}&=(\rho_m+p_m)u^{\mu}u^{\nu}+p_mg^{\mu\nu},
\end{align}
where $\rho_r$ and $p_r$ are the EF radiation energy density and pressure respectively, similarly, $\rho_m$ and $p_m$ are the EF matter energy density and pressure respectively, and $u^{\mu}$ is the 4-velocity in the EF. By projecting the matter and radiation conservation equations along the 4-velocity, we obtain the following modified conservation equations
\begin{align}
u^\mu\nabla_\mu\rho_r+(\rho_r+p_r)\nabla_\mu u^\mu&=-Q_ru^\mu\phi_{,\mu}, \\
u^\mu\nabla_\mu\rho_m+(\rho_m+p_m)\nabla_\mu u^\mu&=-Q_mu^\mu\phi_{,\mu}.
\end{align}
These modified conservation equations show that energy is transferred from the scalar field, depicted by the term projecting the field gradient along the 4-velocity. 
We also define a perfect fluid energy-momentum tensor for both radiation and matter in the RF and JF respectively
\begin{align}
\tilde{T}^{\mu\nu}_{(r)}&=(\tilde{\rho}_r+\tilde{p}_r)\tilde{u}^\mu\tilde{u}^\nu+\tilde{p}_r\tilde{g}^{\mu\nu}_{(r)}\label{TRF},\\
\tilde{T}^{\mu\nu}_{(m)}&=(\tilde{\rho}_m+\tilde{p}_m)\tilde{u}^\mu\tilde{u}^\nu+\tilde{p}_m\tilde{g}^{\mu\nu}_{(m)}.
\end{align}
From now on, we will be considering a time-dependent scalar field, and a zero curvature Friedmann-Robertson-Walker (FRW) EF metric, given by $ds^2=g_{\mu\nu}dx^\mu dx^\nu=-dt^2+a^2(t)\delta_{ij}dx^idx^j$, where $a(t)$ is the expansion scale factor. Following \cite{Barrow}, we introduce an electromagnetic parametrisation, 
\begin{equation}
\eta\equiv\frac{\mathcal{\tilde{L}}_{EM}}{\tilde{\rho}_r},
\end{equation}
where $\eta$ could be positive or negative and have a modulus between 0 and $\approx1$. We should note that in \cite{Barrow} they associate the $\eta$ parameter with matter energy density and not with radiation energy density. Also, by setting $\eta=1$ in the introduced electromagnetic parametrisation, and considering the special case $C_r(\phi)=C_m(\phi)=1$ together with $D_r(\phi)=D_m(\phi)=0$, we recover all the relevant equations given in \cite{MM}, in which a statistical microscopic approach was employed. Using (\ref{density_relation}) we obtain a relationship between the energy density in the RF and that in the EF as follows
\begin{equation}
\tilde{\rho}_r=\frac{\rho_r}{C_r^2}\sqrt{1+\frac{D_r}{C_r}\phi_{,\mu}\phi^{,\mu}}.
\end{equation}
Hence, we can then write the radiation coupling, $Q_r$, in terms of this $\eta$-parametrisation as follows
\begin{equation}\label{Qrnew}
Q_r=\frac{C_r^{\prime}}{2C_r}T_{(r)}+\frac{D_r^{\prime}}{2C_r}\phi_{,\mu}\phi_{,\nu}T^{\mu\nu}_{(r)}+\frac{h^{\prime}}{h}\left[1+\frac{D_r}{C_r}\phi_{,\mu}\phi^{,\mu}\right]\eta\rho_r-\nabla_\mu\left[\frac{D_r}{C_r}\phi_{,\nu}T^{\mu\nu}_{(r)}\right].
\end{equation}
For FRW cosmology, we find that the Klein-Gordon equation, (\ref{klein-gordon}), together with the conservation equations, (\ref{matter_cons}) and (\ref{radiation_cons}), reduce to the following 
\begin{align}
\ddot{\phi}+3H\dot{\phi}+V^{\prime}&=Q_m+Q_r\label{KG},\\
\dot{\rho}_m+3H(\rho_m+p_m)&=-Q_m\dot{\phi}\label{matter},\\
\dot{\rho}_r+3H(\rho_r+p_r)&=-Q_r\dot{\phi}\label{radiation},
\end{align}
where $H=\dot{a}/a$ is the Hubble parameter and dot represents an EF time derivative. 
By using (\ref{KG}), (\ref{matter}), and (\ref{radiation}), we can rewrite (\ref{Qmnew}) and (\ref{Qrnew}) as follows 
\begin{align}
Q_m&=\frac{A_r}{A_rA_m-D_rD_m\rho_r\rho_m}\left[B_m-\frac{D_mB_r}{A_r}\rho_m\right],\\
Q_r&=\frac{A_m}{A_rA_m-D_rD_m\rho_r\rho_m}\left[B_r-\frac{D_rB_m}{A_m}\rho_r\right],
\end{align}
where
\begin{align}
A_r&=C_r+D_r(\rho_r-\dot{\phi}^2),\\
A_m&=C_m+D_m(\rho_m-\dot{\phi}^2),
\end{align}
\begin{align}
B_r&=\frac{1}{2}C_r^{\prime}[3w_r-1]\rho_r-\frac{1}{2}D_r^{\prime}\dot{\phi}^2\rho_r+\frac{h^{\prime}}{h}[C_r-D_r\dot{\phi}^2]\eta\rho_r+D_r\rho_r\left[\frac{C_r^{\prime}}{C_r}\dot{\phi}^2+V^{\prime}+3H\dot{\phi}\left(1+w_r\right)\right],\\
B_m&=\frac{1}{2}C_m^{\prime}[3w_m-1]\rho_m-\frac{1}{2}D_m^{\prime}\dot{\phi}^2\rho_m+D_m\rho_m\left[\frac{C_m^{\prime}}{C_m}\dot{\phi}^2+V^{\prime}+3H\dot{\phi}\left(1+w_m\right)\right].
\end{align}
Using (\ref{w}), it follows that for pressureless matter, the equation of state parameter in both frames is still zero, although according to (\ref{w}), the equation of state parameter for radiation in the EF is now modified to \cite{Carsten,Jack}
\begin{equation}\label{EOS}
w_r=\frac{1}{3}\left(1-\frac{D_r}{C_r}\dot{\phi}^2\right),
\end{equation} 
where we have used the fact that in the RF, $\tilde{w}_r=1/3$. We find that the exact solutions for (\ref{matter}) and (\ref{radiation}) are the following
\begin{align}
\rho_m &\propto\frac{C_m^2}{Y_m}\left(aC_m^{\frac{1}{2}}\right)^{-3},\\
\rho_r &\propto\frac{C_r^2}{h^\eta Y_r}\left(a C_r^{\frac{1}{2}}\right)^{-4} \label{rad_solution},
\end{align}
where we define $Y_m^2=1-(D_m/C_m)\dot{\phi}^2$, $Y_r^2=1-(D_r/C_r)\dot{\phi}^2$ and assume constant $\eta$ in (\ref{rad_solution}). As expected, we find that $\tilde{\rho}_m\propto\tilde{a}_{(m)}^{-3}$ where $\tilde{a}_{(m)}$ is the scale factor in the JF, and that $\tilde{\rho}_r\propto h^{-\eta}\tilde{a}_{(r)}^{-4}$ where $\tilde{a}_{(r)}$ is the RF scale factor. 
\par
We now give the Friedmann equations in the EF obtained from the Einstein Field Equations (\ref{EFE})
\begin{align}
H^2&=\frac{1}{3}\left(\rho_m+\rho_r+\rho_\phi\right)\label{Friedmann},\\
\dot{H}&=-\frac{1}{6}\left[3\left(\rho_m+\dot{\phi}^2\right)+\rho_r\left(4-\frac{D_r}{C_r}\dot{\phi}^2\right)\right],
\end{align}
where we have also used (\ref{EOS}), and defined the energy-momentum tensor of the scalar field as that of a perfect fluid with $\rho_\phi=(1/2)\dot{\phi}^2+V(\phi)$ and $p_\phi=(1/2)\dot{\phi}^2-V(\phi)$.
\par
We admit that the scalar field characterizing the disformal coupling is also responsible for the current acceleration of the Universe, i.e., it is the dark energy. 
In order to compare our model's dark energy equation of state parameter to type Ia supernova data set \cite{Suzuki}, we need to derive an effective equation of state parameter, $w_{\text{eff}}$, following \cite{Khoury,Jack}. Experimental constraints on dark energy assume a non-interacting dark sector. On assuming that dark energy is given by a non-interacting perfect fluid, described by its equation of state parameter, $w_{\text{eff}}$, we can write the energy conservation equation as follows
\begin{equation}\label{eff}
\dot{\rho}^{\text{eff}}_{\text{DE}}=-3H(1+w_{\text{eff}})\rho^{\text{eff}}_{\text{DE}}.
\end{equation}
By our dark sector assumption, dark matter is also assumed to be non-interacting, hence we can write the following Friedmann equation
\begin{equation}\label{Fried}
H^2=\frac{1}{3}\left(a^{-4}\rho_{0,r}+a^{-3}\rho_{0,m}+\rho_{\text{DE}}^{\text{eff}}\right).
\end{equation}
By comparing (\ref{Fried}) and (\ref{Friedmann}), we get the energy density of the effective dark energy fluid
\begin{equation}\label{DE}
\rho^{\text{eff}}_{\text{DE}}=\rho_m+\rho_r+\rho_{\phi}-a^{-4}\rho_{0,r}-a^{-3}\rho_{0,m}.
\end{equation}
By taking the EF time derivative of (\ref{DE}), substituting the Klein-Gordon equation, (\ref{KG}), the matter conservation equation, (\ref{matter}), the radiation conservation equation, (\ref{radiation}), and comparing the final equation with (\ref{eff}), we get the effective equation of state parameter as follows
\begin{equation}
w_{\text{eff}}=\frac{p_{\phi}+\rho_r\left(w_r-\frac{1}{3}a^{-4}\frac{\rho_{0,r}}{\rho_r}\right)}{\rho_{\text{DE}}^{\text{eff}}},
\end{equation}
where $w_r$ is the equation of state parameter for radiation in the EF given by (\ref{EOS}). We should mention that in the absence of radiation, matter and electromagnetic couplings, $w_{\text{eff}}$ reduces to the usual equation of state parameter for a quintessence scalar field, $w_{\phi}=p_{\phi}/\rho_{\phi}$. 

%%%%%%%%%%%%%%
\section{Observational Constraints}\label{sec:constraints}
%%%%%%%%%%%%%%
We will only consider a choice of parameters for our models such that we agree with the measured cosmological, astrophysical, geochemical and laboratory parameters. Since we now have identified the fine-structure coupling  $\alpha\propto Z\;h(\phi)^{-1}$, we can also define the temporal variation of $\alpha$, denoted by $\dot{\alpha}/\alpha$, where dot refers to the temporal derivative. As we are interested in solving our equations in a spatially-flat, homogeneous and isotropic Friedmann gravitational metric, and also a time-dependent scalar field, we here give the temporal variation of $\alpha$ in the mentioned setting. In this scenario, the coupling function $Z$, defined in (\ref{Zdefinition}), reduces to
\begin{equation}
Z=\left(\frac{1-\frac{D_r}{C_r}\dot{\phi}^2}{1-\frac{D_m}{C_m}\dot{\phi}^2}\right)^{\frac{1}{2}}~.
\end{equation}
Furthermore, by considering the electromagnetic coupling, $h$, together with all the conformal and disformal couplings to be a function of $\phi$ only, it follows that $Z(\phi,\dot{\phi})$. We then arrive to the equation for the temporal variation of the fine-structure coupling  
\begin{equation}\label{alpha}
\frac{\dot{\alpha}}{\alpha}=\frac{1}{Z}\left(\frac{\partial Z}{\partial\phi}\dot{\phi}+\frac{\partial Z}{\partial\dot{\phi}}\ddot{\phi}\right)-\frac{1}{h}\frac{dh}{d\phi}\dot{\phi}.
\end{equation} 
Also, the redshift evolution of the fine-structure coupling is specified by the quantity\footnote{Note that the redshift is frame-invariant, see \cite{Carsten} and \cite{Brax}.} 
\begin{equation}
\frac{\Delta \alpha}{\alpha}(z) \equiv \frac{\alpha(z) - \alpha(z=0)}{\alpha(z=0)}=\frac{h(\phi_0)Z(z)}{h(\phi(z))Z_0} - 1,
\end{equation}
where $\phi_0$ is the field value today and $Z_0$ is the value of $Z$ evaluated today. 
We recover the usual forms of the temporal variation and evolution of the fine-structure coupling, {in Refs.~\cite{MM,Dam1,Dam2,Dam3,Dam4,Nunes1,Olive,Nunes2,Uzan}, 
 in the absence of disformal couplings}. In order to choose our parameters will use:
\begin{enumerate}
\item the Union2.1 SNe Ia data set \cite{Suzuki} and the Planck collaboration results \cite{P2}, such that the present time measurements of our Universe agree with our final time boundary conditions. These results are summarized in Table \ref{table1}. Since we are using a spatially-flat FRW EF metric, it follows that $\Omega_{0,\phi}\approx0.7$. 
\begin{table}[t]
\begin{center}
\begin{tabular}{ c c c c c}
 \hline
\hline
 Parameter &  Estimated value & Ref. \\ 
\hline
$w_{0,\phi}$ & $-1.006\pm0.045$  & \cite{P2}\\ 

 $H_0$ & $(67.8\pm0.9)\;\text{km}\;\text{s}^{-1}\text{Mpc}^{-1}$ & \cite{P2}\\ 

$\Omega_{0,m}$ & $0.308\pm0.012$  & \cite{P2}\\ 
 \hline
\hline
\end{tabular}
\caption{\label{table1} Listed are, respectively, the cosmological parameter, its estimated value, and the original reference.}
\end{center}
\end{table}

\item the currently, most stringent atomic clock (AC) constraint on the present temporal variation of $\alpha$ \cite{M}
\begin{equation}\label{rosenband}
\left.\frac{\dot{\alpha}}{\alpha}\right\vert_0=(-1.6\pm 2.3)\times 10^{-17}\;\text{yr}^{-1},
\end{equation}

\item an Oklo natural reactor constraint, in which self-sustained natural fission reactions took place at $\sim$ 2 Gyr ago $(z\simeq0.16)$ \cite{Fujii,Davis}
\begin{equation}
\frac{|\Delta\alpha|}{\alpha}<1.1\times10^{-8},
%\frac{\Delta\alpha}{\alpha}=(-0.8\pm1)\times10^{-8}
\end{equation}

\item the ${^{187}}\text{Re}$ meteorite constraint over the age of the solar system $\sim$ 4.6 Gyr $(z\simeq0.43)$ \cite{O} 
\begin{equation}
\frac{\Delta\alpha}{\alpha}=(-8\pm8)\times10^{-7},
\end{equation}

\item astrophysical data, including the 11 recently measured data set which contains the results of the Ultraviolet and Visual Echelle Spectrograph (UVES) \cite{J,K}. We list these measurements in Table \ref{table2}, and plot the corresponding data points for the redshift evolution of $\Delta\alpha/\alpha$ in the figures of Section \ref{sec:analysis}. We also use the large data set from Keck telescope and ESO's Very Large Telescope (VLT) surveys carried out by Webb \textit{et al} \cite{E}. By assuming that $\Delta\alpha/\alpha$ values are described by a simple weighted mean, it is found that \cite{Murphy04} $(\Delta\alpha/\alpha)_\text{w}=(-0.57\pm0.11)\times10^{-5}$ for the Keck quasi-stellar object (QSO) observations; while for the VLT quasar spectra observations, it is found that \cite{King} $(\Delta\alpha/\alpha)_\text{w}=(0.208\pm0.124)\times10^{-5}$. These results differ from one another at the $\sim4.7\sigma$ level, suggesting a dipole-like variation in $\alpha$. The binned many-multiplet (MM) VLT+Keck combined sample is also used in our $\Delta\alpha/\alpha$ redshift evolution plots, in which statistical errors for certain points have been increased prior to binning as reported in King \textit{et al} \cite{King}. We also use the weighted mean of the 21 Si IV doublets, $(\Delta\alpha/\alpha)_\text{w}=(-0.5\pm1.3)\times10^{-5}$, reported in Murphy \textit{et al} \cite{AD} using the alkali-doublet (AD) method, together with two other results; $\Delta\alpha/\alpha=(-0.10\pm0.22)\times10^{-5}$  at $z=0.25$ and $\Delta\alpha/\alpha=(-0.08\pm0.27)\times10^{-5}$  at $z=0.68$; in which HI 21 cm absorption lines were used \cite{N}. Other consistent results can be found in \cite{Rahmani}, and a detailed review on the constraints on $\Delta\alpha/\alpha$ can be found in \cite{Uzan,Chiba}.
\begin{table}[t]
\begin{center}
\begin{tabular}{ l c c c c} 
 \hline
\hline
 Object &  $z$ & $(\Delta\alpha/\alpha)\times10^6$ & Spectrograph & Ref. \\ 
\hline
Three sources & 1.08 & $4.3\pm3.4$ & HIRES & \cite{F}  \\ 
HS1549+1919 & 1.14 & $-7.5\pm5.5$ & UVES/HIRES/HDS & \cite{K}  \\ 
HE0515-4414 & 1.15 & $-0.1\pm1.8$ & UVES & \cite{G}  \\ 
HE0515-4414 & 1.15 & $0.5\pm2.4$ & HARPS/UVES & \cite{H}  \\ 
HS1549+1919 & 1.34 & $-0.7\pm6.6$ & UVES/HIRES/HDS & \cite{K}  \\ 
HE0001-2340 & 1.58 & $-1.5\pm2.6$ & UVES & \cite{I}  \\ 
HE1104-1805A & 1.66 & $-4.7\pm5.3$ & HIRES & \cite{F}  \\ 
HE2217-2818 & 1.69 & $1.3\pm2.6$ & UVES & \cite{J}  \\ 
HS1946+7658 & 1.74 & $-7.9\pm6.2$ & HIRES & \cite{F}  \\ 
HS1549+1919 & 1.80 & $-6.4\pm7.2$ & UVES/HIRES/HDS & \cite{K}  \\ 
Q1101-264 & 1.84 & $5.7\pm2.7$ & UVES & \cite{G}  \\ 
 \hline
\hline
\end{tabular}
\end{center}
\caption{\label{table2} Listed are, respectively, the object along each line of sight, the redshift of the absorber, the measurements of $\Delta\alpha/\alpha$, the spectrograph, and the original reference. The first measurement is the weighted average from eight absorbers in the redshift range $0.73<z<1.53$ along the lines of sight of three quasars reported in \cite{F}.}
\end{table}

\item and the cosmic microwave background (CMB) radiation constraint $(z\simeq10^3)$ reported in \cite{P1}
\begin{equation}
\frac{\Delta\alpha}{\alpha}=(3.6\pm3.7)\times10^{-3}.
\end{equation} 

\end{enumerate}
The models presented in Section  \ref{sec:analysis} satisfy the above constraints, even though, there are combinations of parameters that lead to cases which do not respect the Oklo bound. For example, some models in Fig. \ref{fig:disformal_EM} do not pass the Oklo bound.

\section{Analysis}\label{sec:analysis}
\hspace{1cm}In general, we will consider these specific forms of couplings and scalar field potential:
\begin{align}
C_i(\phi)&=\beta_i e^{x_i\phi},\\
D_i(\phi)&=M_i^{-4}e^{y_i\phi},\\
h(\phi)&=1-\zeta(\phi-\phi_0)\label{coupling},\\
V(\phi)&=M_V^4e^{-\lambda\phi}.
\end{align} 
The introduced mass scales, $M_i$ and $M_V$, are tuned in order to obtain the correct cosmological parameters as listed in Table \ref{table1} together with an agreement with the temporal variation of the fine-structure coupling (\ref{rosenband}). Typically we find that $M_i\sim M_V \sim$ meV. In our models, we will only consider disformally coupled radiation, and hence we set $C_r(\phi)=1$ for all the models under consideration. On the other hand, whenever we consider a scalar field dependent conformal matter coupling, we also tune the dimensionless parameter, $\beta_m$, together with the other mass scales in order to agree with the measured cosmological parameters and temporal variation of the fine-structure coupling stringent constraint. Following the symmetry breaking argument in \cite{Dvali} for a slowly evolving time dependent scalar field, we only consider a linear electromagnetic coupling function. We are constrained on the magnitude of the dimensionless electromagnetic coupling parameter, $\zeta$, from local tests of the equivalence principle \cite{Uzan}
\begin{equation}
|\zeta_{\text{local}}|<10^{-3}
\end{equation}
Another constraint on $\zeta$ was obtained in \cite{Calabrese} using the CMB and large-scale structure data in combination with direct measurements of the expansion of the Universe, and more recently, another tighter constraint was obtained in \cite{Pinho} using more recent data, in which it was found that $|\zeta|<5\times10^{-6}$. 
\par
We will now consider several models with different parameters. From our numerical results, we found that a change in the magnitude of $|\eta|$ between $0$ and $1$ has negligible effect on the results, so we set $\eta=1$ in all the models. We summarize the different parameter values for each specific model in Table \ref{table3}, where we have neglected the parameters, $y_m$ and $y_r$, from the table as these parameters are both set to zero for the models shown in Fig. \ref{fig:disformal_EM}-\ref{fig:conformal_disformal_EM}, although we do discuss a model with exponential disformal couplings in Section \ref{sec:disformal}.
\begin{table}[t]
\begin{center}
\begin{tabular}{ c c c c c c c c c} 
 \hline
\hline
 Fig. &  $M_r$  & $M_m$ & $\beta_m$ & $x_m$ & $|\zeta|$ & $M_V$ & $\lambda$ & $\left.(\dot{\alpha}/\alpha)\right\vert_0\times10^{17}$\\ 
\hline
\ref{fig:disformal_EM} & $\sim$ meV & $\sim$ meV & 1 & 0 & $<5\times10^{-6}$ & 2.69 meV & 0.45 & $-2.14\sim-1.62$  \\ 
\ref{fig:conformal_disformal} & $\sim$ meV & 15 meV & 8 & 0.14 & 0 & 2.55 meV & 0.45 & $-2.41\sim0.70$ \\ 
\ref{fig:conformal_disformal_EM} & $\sim$ meV & 15 meV & 8 & 0.14 & $<5\times10^{-6}$ & 2.55 meV & 0.45 & $-2.10\sim-1.24$ \\ 
 \hline
\hline
\end{tabular}
\end{center}
\caption{\label{table3} Listed are, respectively, the figure reference, the parameter values for each specific model, and the range of $\left.(\dot{\alpha}/\alpha)\right\vert_0$ for the range of parameter values considered in each figure.}
\end{table}
The data points shown in the evolution with redshift of $\Delta\alpha/\alpha$ in Fig. \ref{fig:disformal_EM}-\ref{fig:conformal_disformal_EM} were taken from  Section~\ref{sec:constraints}, as described therein. The respective data sets of the data points are listed in the legend of each plot, where UVES+HIRES+HDS+HARPS refer to the tabulated data points in Table \ref{table2}.  The constraints on $w_{\text{eff}}$ used in the plots depicting the evolution with redshift of the theoretical $w_{\text{eff}},w_{\phi}$ in Fig. \ref{fig:disformal_EM}-\ref{fig:conformal_disformal_EM} were taken from \cite{Suzuki}, where the results were obtained assuming a flat Universe for the joint data set of SNe, BAO, CMB, and $H_0$, with (dark/orange) and without (light/yellow) SN systematics. The dotted and dot-dashed lines in these plots depict the central value of $w_{\text{eff}}$ constraints with and without SN systematics in each redshift bin respectively. Although we start integrating our equations from $z=10^3$, at which we set $\phi_{\text{ini}}=1.5\;\text{M}_{\text{Pl}}$  as our initial condition, we restrict our plots in Fig. \ref{fig:disformal_EM}-\ref{fig:conformal_disformal_EM} to the redshift range where observational data is mostly concentrated.

\subsection{Disformal couplings}\label{sec:disformal}
\begin{figure}[h!]
{\centering
\begin{tabular*}{\textwidth}{@{}cc@{}}
  \includegraphics[width=0.47\textwidth]{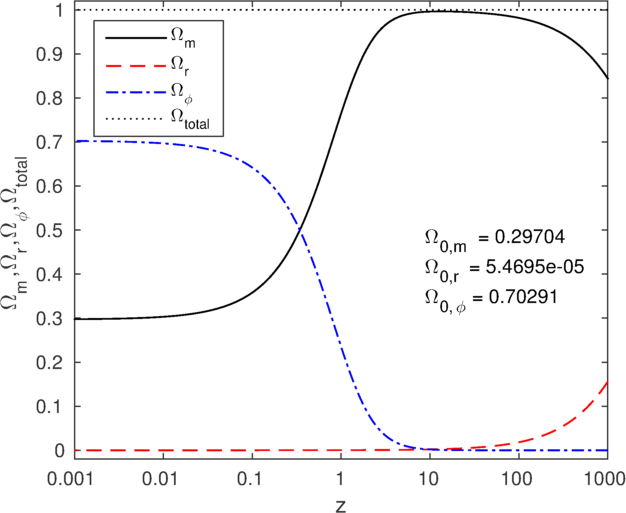} &
  \includegraphics[width=0.5\textwidth]{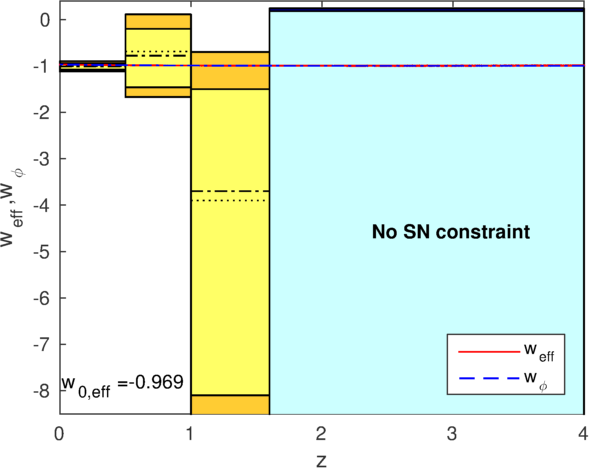}
  \end{tabular*}\par}
{\centering
\begin{tabular*}{\textwidth}{@{}c@{}}
  \includegraphics[width=1\textwidth]{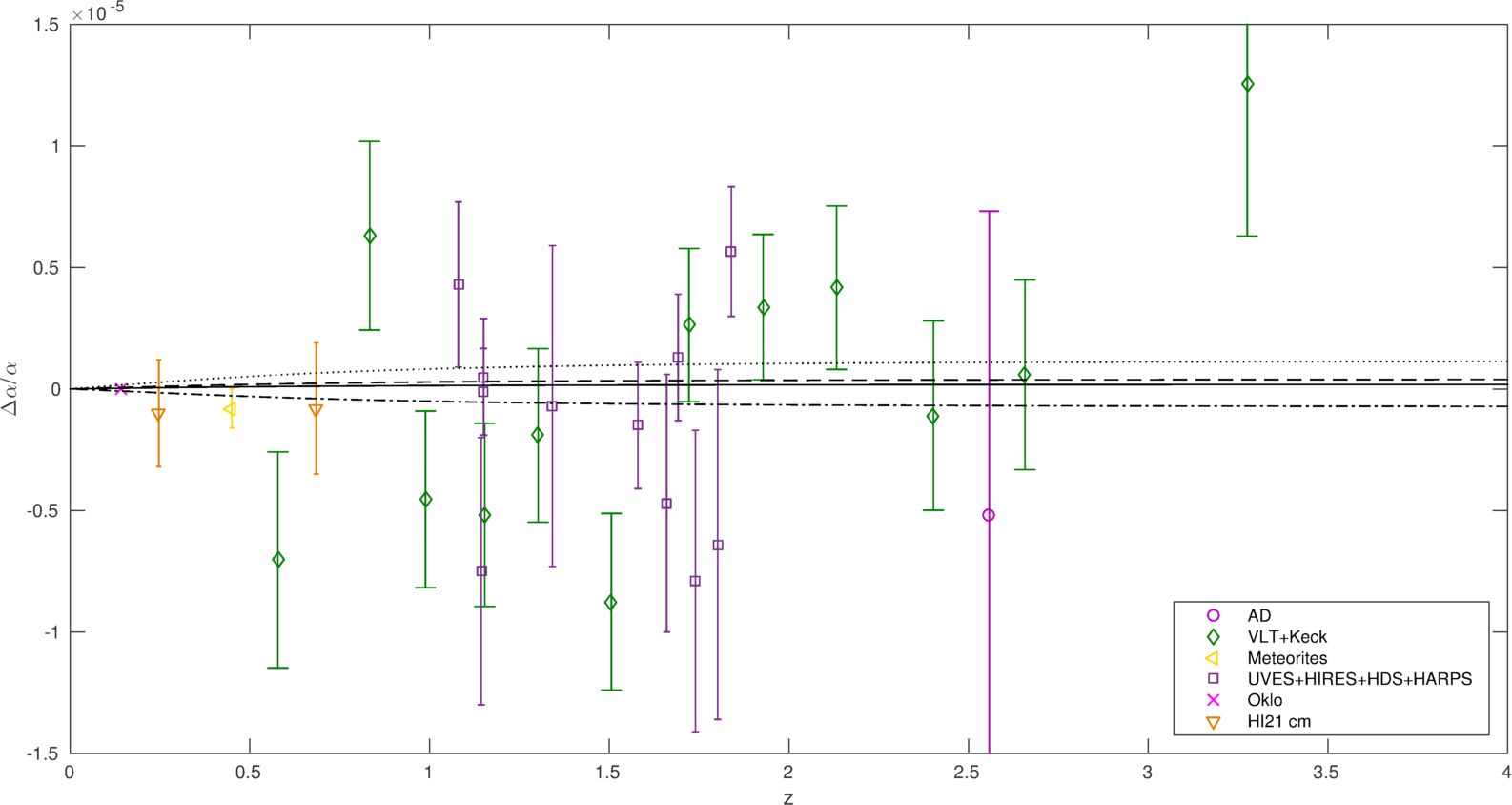}
 \end{tabular*}
\caption{Redshift evolution of $\Omega_i$'s, $w_{\text{eff}}$, $w_{\phi}$ and $\Delta\alpha/\alpha$ when considering disformally coupled matter and radiation with and without an electromagnetic coupling. The solid line is the purely disformal case, whereas the dot-dashed, dashed, and dotted lines correspond to $\zeta=-4.9\times10^{-6},\;1\times10^{-6},\;4.9\times10^{-6}$ respectively. The other model parameters are summarized in Table \ref{table3}.} 
\label{fig:disformal_EM}\par}
\end{figure}

\hspace{1cm}In our first model, we consider disformally coupled radiation and matter in the absence of matter and radiation conformal couplings and set $\zeta=0$ in order to get a purely disformal case, without the addition of an electromagnetic coupling. This purely disformal case is an interesting model to consider, since it still predicts a non-zero variation of the fine-structure coupling, without the need of an electromagnetic coupling. The redshift evolution of $\Delta\alpha/\alpha$ is shown in Fig. \ref{fig:disformal_EM}, in which we consider constant matter and radiation disformal couplings. We have also considered a purely disformal case with exponential disformal couplings, by setting $M_m=M_r=100$ meV, $y_m=23.5$ and $y_r=1$, together with $M_V=4.75$ meV and $\lambda=2$. The latter two parameters were changed from those presented in Table \ref{table3} in order to be in agreement with the current cosmological parameters, whereas the exponential disformal coupling parameters were chosen such that the calculated temporal variation of the fine-structure coupling lies within the bounds of the estimated value measured by AC. Since the redshift evolutions of $\Delta\alpha/\alpha$, $\Omega_i$'s, $w_\text{eff}$, and $w_{\phi}$, for constant disformal couplings and exponential disformal couplings are indistinguishable, we only show the constant disformal couplings case in Fig. \ref{fig:disformal_EM}.

\subsection{Conformal and disformal couplings}\label{sec:conformal_disformal}
\begin{figure}[h!]
{\centering
\begin{tabular*}{\textwidth}{@{}cc@{}}
  \includegraphics[width=0.47\textwidth]{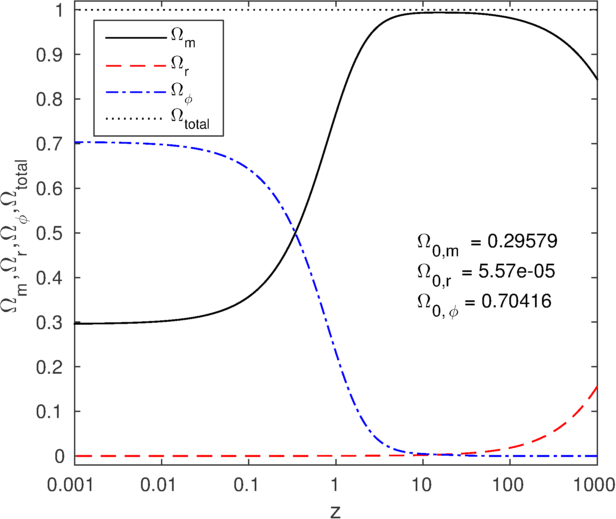} &
  \includegraphics[width=0.5\textwidth]{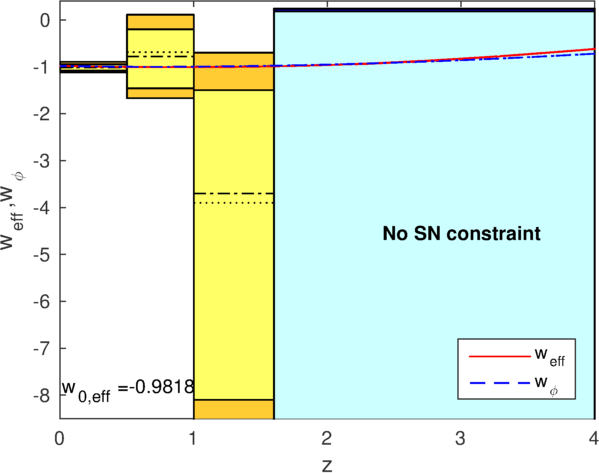}
  \end{tabular*}\par}
{\centering
\begin{tabular*}{\textwidth}{@{}c@{}}
  \includegraphics[width=1\textwidth]{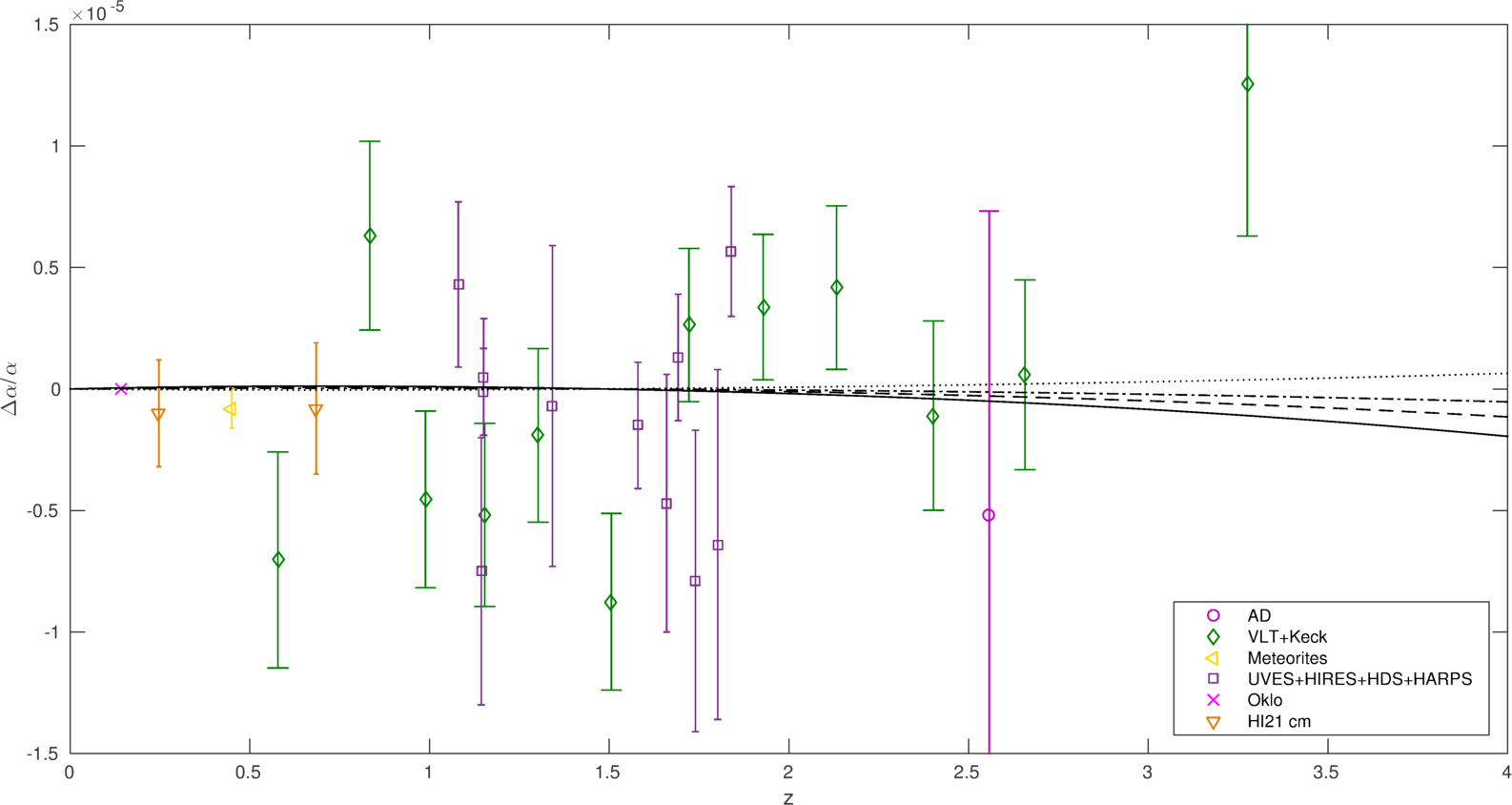}
 \end{tabular*}
\caption{Redshift evolution of $\Omega_i$'s, $w_{\text{eff}}$, $w_{\phi}$ and $\Delta\alpha/\alpha$ when considering conformally and disformally coupled matter together with disformally coupled radiation in the absence of an electromagnetic coupling. The solid, dashed, dot-dashed, and dotted lines shown in the redshift evolution of $\Delta\alpha/\alpha$ plot correspond to $M_r=24.91, 25.45, 25.91, 26.91$ meV respectively. The other parameters are summarized in Table \ref{table3}.} 
\label{fig:conformal_disformal}\par}
\end{figure}

\hspace{1cm}We now analyse the case in which we only consider matter to be conformally and disformally coupled together with disformally coupled radiation. An exponential matter conformal coupling together with constant disformal matter and radiation couplings are considered, as summarized in Table \ref{table3}. The range of values for the constant disformal couplings are chosen such that the predicted current temporal variation in the fine-structure constant lies within the limits of the constraint given by (\ref{rosenband}), as these are the model parameters which directly affect the evolution of $\alpha$. We also restrict the values of the other model parameters, mainly the scalar field potential and the conformal matter coupling parameters in order to be in agreement with the current cosmological values listed in Table \ref{table1}. Four different constant radiation energy scales are used in this model and shown in Fig. \ref{fig:conformal_disformal}. We demonstrate that one can also get a non-zero value for the variation of $\alpha$, without the introduction of an electromagnetic coupling. This shows that we can still explain observational evidence for such a variation in the fine-structure coupling, without the need of an extra electromagnetic coupling. In these models $|\Delta\alpha/\alpha|$ at the CMB era is predicted to be $\mathcal{O}(10^{-8}-10^{-7})$ which is well within the current bounds.     

\subsection{Disformal and electromagnetic couplings}\label{sec:disformal_EM}
\hspace{1cm}In this case, we consider matter and radiation to be disformally coupled together with an electromagnetic coupling. We again choose our model parameters so that we are in agreement with the data mentioned in Section ~\ref{sec:constraints}. In Fig. \ref{fig:disformal_EM} we only consider constant matter and radiation couplings, as described in Table \ref{table3}. These disformal mass scales together with the non-zero $\zeta$ determine the evolution of the fine-structure coupling. We consider three different choices of these parameters, each giving a different redshift evolution of $\alpha$, although they all predict the same cosmology as long as the other independent parameters are left the same. The range of $|\Delta\alpha/\alpha|$ for these models at the CMB redshift is $\mathcal{O}(10^{-8}-10^{-6})$ which lies within the current CMB constraint. In Fig. \ref{fig:disformal_EM} we also plot the purely disformal case, discussed in Section \ref{sec:disformal}, in order to see the effect of the electromagnetic coupling on disformally coupled matter and radiation. We can clearly see that an electromagnetic coupling enhances the cosmological evolution of $\Delta\alpha/\alpha$.   
        
\subsection{Disformal, conformal and electromagnetic couplings}
\begin{figure}[h!]
{\centering
\begin{tabular*}{\textwidth}{@{}cc@{}}
  \includegraphics[width=0.47\textwidth]{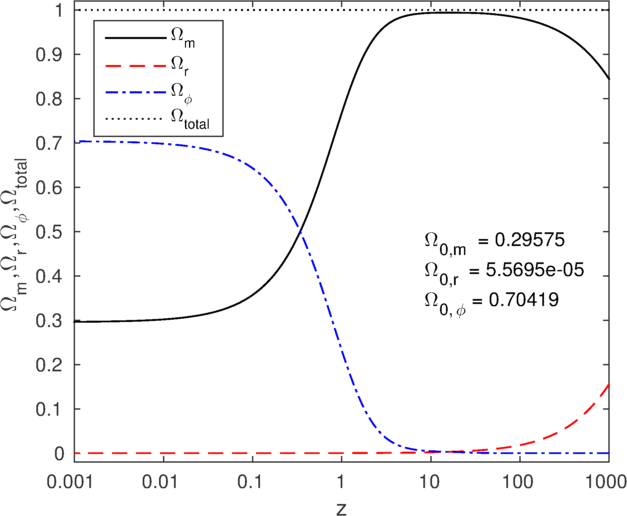} &
  \includegraphics[width=0.5\textwidth]{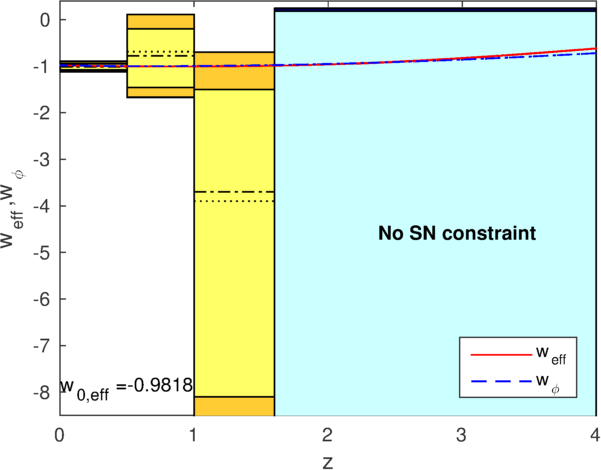}
  \end{tabular*}\par}
{\centering
\begin{tabular*}{\textwidth}{@{}c@{}}
  \includegraphics[width=1.01\textwidth]{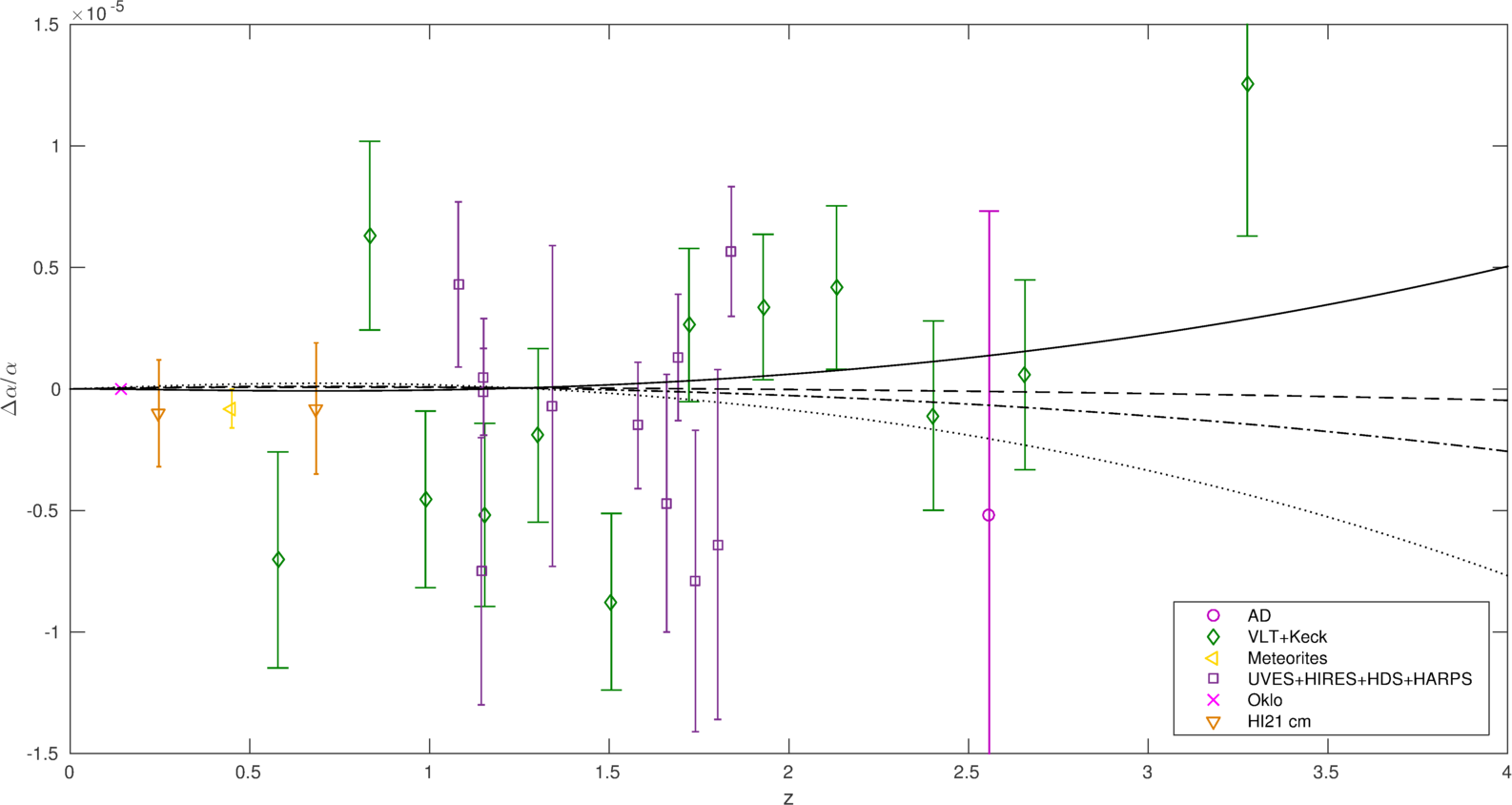}
 \end{tabular*}
\caption{Redshift evolution of $\Omega_i$'s, $w_{\text{eff}}$, $w_{\phi}$ and $\Delta\alpha/\alpha$ when considering conformally and disformally coupled matter, and disformally coupled radiation in the presence of an electromagnetic coupling. The solid, dashed, dot-dashed and dotted lines correspond to $\zeta=-4.9\times10^{-6},\;-1\times10^{-6},\;1\times10^{-6},\;4.9\times10^{-6}$ respectively. The other model parameters are summarized in Table \ref{table3}.} 
\label{fig:conformal_disformal_EM}\par}
\end{figure}

\hspace{1cm}We now combine the previous cases together, where we are considering conformally and disformally coupled matter, and disformally coupled radiation, in the presence of an electromagnetic coupling. We choose a linear electromagnetic coupling as in (\ref{coupling}) and an exponential conformal matter coupling. As in the previous cases, we choose our model parameters so that we are in agreement with the data mentioned in Section ~\ref{sec:constraints}. In Fig. \ref{fig:conformal_disformal_EM} we show different theoretical predictions of the variation of $\alpha$, when considering four different $\zeta$ and $M_r$ parameter choices as summarized in Table \ref{table3}. In general, at least for our chosen parameters, we can conclude that the introduction of an electromagnetic coupling slightly enhances the variation of the fine-structure constant, still in agreement with current observational data. Indeed, for these models, $|\Delta\alpha/\alpha|$ at the CMB era is $\mathcal{O}(10^{-7}-10^{-6})$, still within the CMB constraint. 

%%%%%%%%
\section{Conclusions}
%%%%%%%%
\hspace{1cm}We have introduced the idea of disformally related metrics in order to explain the reported variations in the fine-structure constant. Our main conclusion is that we can naturally explain a non-zero variation in $\alpha$ by only considering disformally coupled radiation and matter. The introduction of an electromagnetic coupling has been exhaustively discussed in the literature and such models also predict a varying fine-structure constant. We have also considered the effect of this electromagnetic coupling in some of our models, and we can say that this coupling enhances the variational evolution of $\alpha$. We made sure that our chosen model parameters predict the current cosmological parameters in agreement with the experimental values together with an agreement with the reported bounds on variations in the fine-structure constant. From our numerical results, we found that when the present temporal variation of the fine-structure constant constraint is satisfied, the other geochemical and astrophysical constraints are also satisfied. We should still mention that the Oklo bound is not always satisfied by the chosen model parameters, although we do find some specific models which predict a variation in $\alpha$ that lies well within the Oklo constraint. We have also considered the weak constraint from the CMB, which is well satisfied in all our models.  
\par
High-resolution ultra-stable spectrographs are expected to increase the accuracy of the currently reported spectroscopic measurements. Such next generation spectrographs include PEPSI at the Large Binocular Telescope (LBT) in Arizona, ESPRESSO at the VLT \cite{ESPRESSO}, and ELT-HIRES at the European Extremely Large Telescope (E-ELT) \cite{ELT1,ELT2}.  Consistency tests like the Sandage-Loeb cosmological redshift drift test could also be carried out in such facilities \cite{Sandage1,Sandage2,Sandage3,Sandage4}. The Atacama Large Millimeter/submillimeter Array (ALMA) \cite{ALMA1,ALMA2} is also expected to increase the sensitivity to detect radio continuum sources. Laboratory measurements with molecular and nuclear clocks are also expected to increase their sensitivity to as high as $10^{-21}\;\text{yr}^{-1}$ \cite{Martins}. Other observational constraints coming from compact objects, such as white-dwarfs \cite{white_dwarf1,white_dwarf2}, are equally being explored.

%%%%%%%%%%%%%%%%%%%%%%%%%%%%%%%%%%%%%%%%%%%%%%%
%%%%%%%%%%%%%%%%%%%%%%%%%%%%%%%%%%%%%%%%%%%%%%%
\begin{acknowledgments}
CvdB is grateful to Clare Burrage and Jack Morrice for fruitful discussions on disformal electrodynamics. The work of CvdB is supported by the Lancaster- Manchester-Sheffield Consortium for Fundamental Physics under STFC Grant No. ST/L000520/1. 
N.J.N was supported by the Funda\c{c}\~{a}o para a Ci\^{e}ncia e Tecnologia 
(FCT) through the grants EXPL/FIS-AST/1608/2013 and UID/FIS/04434/2013. 
\end{acknowledgments}
%%%%%%%%%%%%%%%%%%%%%%%%%%%%%%%%%%%%%%%%%%%%%

\clearpage
\begin{subappendices}
\section*{Appendix A: Disformal Transformations}
\renewcommand{\theequation}{A.\arabic{equation}}
\renewcommand{\thesection}{A}
\refstepcounter{section}
\label{appendix:a}
We here derive some variable transformations between the EF and the tilde frames, the JF and RF. These mappings are also discussed in \cite{Carsten,Zuma1,Zuma2,Minamitsuji,Zuma3,Jack}, although we reproduce the relevant transformations for completeness. To keep the discussion simple, we write the JF metric (\ref{g_m}) and RF metric (\ref{g_r}) as a single tilde metric, which we write as
\begin{equation}\label{tilde_g}
\tilde{g}_{\mu\nu}=Cg_{\mu\nu}+D\phi_{,\mu}\phi_{,\nu},
\end{equation}
and then apply the variable transformations to each frame. From (\ref{tilde_g}), it follows that the inverse metric is given by
\begin{equation}\label{inverse_g}
\tilde{g}^{\mu\nu}=\frac{1}{C}\left(g^{\mu\nu}-\bar{\gamma}^2\partial^\mu\phi\partial^\nu\phi\right),
\end{equation}
where
\begin{equation}\label{gamma_g}
\bar{\gamma}^2=\frac{D}{C+Dg^{\mu\nu}\phi_{,\mu}\phi_{,\nu}}.
\end{equation}
The determinants are related via
\begin{equation}\label{det_g}
\sqrt{\frac{-\tilde{g}}{-g}}=C^2\sqrt{1+\frac{D}{C}\phi_{,\mu}\phi^{,\mu}}.
\end{equation}
We now derive a relationship between the energy-momentum tensors in the EF metric and in the tilde metric (\ref{tilde_g}). Using the definition of the energy-momentum tensor and the chain rule, we obtain the following 
\begin{equation}
T^{\mu\nu}=\frac{2}{\sqrt{-g}}\frac{\delta(\sqrt{-\tilde{g}}\mathcal{\tilde{L}})}{\delta g_{\mu\nu}}=\sqrt{\frac{-\tilde{g}}{-g}}\frac{\delta\tilde{g}_{\alpha\beta}}{\delta g_{\mu\nu}}\left(\frac{2}{\sqrt{-\tilde{g}}}\frac{\delta(\sqrt{-\tilde{g}}\mathcal{\tilde{L}})}{\delta\tilde{g}_{\alpha\beta}}\right)=\sqrt{\frac{-\tilde{g}}{-g}}\frac{\delta\tilde{g}_{\alpha\beta}}{\delta g_{\mu\nu}}\tilde{T}^{\alpha\beta}.
\end{equation}
Hence, together with (\ref{tilde_g})--(\ref{det_g}), the contravariant and mixed energy-momentum tensor relations between the EF metric and tilde metric are as follows
\begin{align}
T^{\mu\nu}&=C^3\sqrt{1+\frac{D}{C}\phi_{,\mu}\phi^{,\mu}}\tilde{T}^{\mu\nu},\\
T^\mu_{\;\;\;\nu}&=C^2\sqrt{1+\frac{D}{C}\phi_{,\mu}\phi^{,\mu}}\left[\delta^\rho_{\;\;\;\nu}-\frac{D\phi_{,\nu}\phi^{,\rho}}{C+D\phi_{,\mu}\phi^{,\mu}}\right]\tilde{T}^\mu_{\;\;\;\rho}\label{transformation}.
\end{align}
On specifying a perfect fluid energy-momentum tensor in both frames, 
\begin{align}
T^{\mu\nu}&=(\rho+p)u^{\mu}u^{\nu}+pg^{\mu\nu},\\
\tilde{T}^{\mu\nu}&=(\tilde{\rho}+\tilde{p})\tilde{u}^\mu\tilde{u}^\nu+\tilde{p}\tilde{g}^{\mu\nu}.
\end{align}
we get the following mappings for the energy density, pressure and the equation of state parameter
\begin{align}
\tilde{\rho}&=\frac{\rho}{C^2}\sqrt{1+\frac{D}{C}\phi_{,\mu}\phi^{,\mu}}\label{density_relation},\\
\tilde{p}&=\frac{p}{C^2\sqrt{1+\frac{D}{C}\phi_{,\mu}\phi^{,\mu}}},\\
\tilde{w}\equiv\frac{\tilde{p}}{\tilde{\rho}}&=\frac{w}{1+\frac{D}{C}\phi_{,\mu}\phi^{,\mu}}\label{w},
\end{align}
when considering a spatially-flat FRW EF metric and a time-dependent scalar field. Using this tilde metric (\ref{tilde_g}), we also derive a relationship between the 4-velocity in the EF, $u^{\mu}$, and that in the tilde frame, $\tilde{u}^\mu$, given by the following equation
\begin{equation}
\tilde{u}^\mu=\frac{u^\mu}{\sqrt{C}\sqrt{1+\frac{D}{C}\phi_{,\mu}\phi^{,\mu}}}.
\end{equation}
\end{subappendices}

\bibliographystyle{JHEP}
\bibliography{mybib}

\providecommand{\href}[2]{#2}\begingroup\raggedright\begin{thebibliography}{10}

\bibitem{Dirac}
P.~A.~M. Dirac, {\it {The Cosmological constants}},  {\em Nature} {\bf 139}
  (1937) 323.

\bibitem{Chodos}
A.~Chodos and S.~L. Detweiler, {\it {Where Has the Fifth-Dimension Gone?}},
  {\em Phys. Rev.} {\bf D21} (1980) 2167.

\bibitem{Wu}
Y.-S. Wu and Z.~Wang, {\it {The Time Variation of Newton's Gravitational
  Constant in Superstring Theories}},  {\em Phys. Rev. Lett.} {\bf 57} (1986)
  1978.

\bibitem{CODATA}
P.~Mohr, B.~Taylor, and D.~Newell, {\it The 2014 codata recommended values of
  the fundamental physical constants (web version 7.0)},  2015.
\newblock This database was developed by J. Baker, M. Douma, and S.
  Kotochigova. Available: http://physics.nist.gov/constants [Thursday,
  27-Aug-2015 10:53:20 EDT]. National Institute of Standards and Technology,
  Gaithersburg, MD 20899.

\bibitem{Iocco}
F.~Iocco, G.~Mangano, G.~Miele, O.~Pisanti, and P.~D. Serpico, {\it {Primordial
  Nucleosynthesis: from precision cosmology to fundamental physics}},  {\em
  Phys. Rept.} {\bf 472} (2009) 1--76,
  [\href{http://arxiv.org/abs/0809.0631}{{\tt arXiv:0809.0631}}].

\bibitem{P1}
{\bf Planck:} Collaboration, P.~A.~R. Ade et~al., {\it {Planck intermediate
  results - XXIV. Constraints on variations in fundamental constants}},  {\em
  Astron. Astrophys.} {\bf 580} (2015) A22,
  [\href{http://arxiv.org/abs/1406.7482}{{\tt arXiv:1406.7482}}].

\bibitem{E}
J.~K. Webb, J.~A. King, M.~T. Murphy, V.~V. Flambaum, R.~F. Carswell, and M.~B.
  Bainbridge, {\it {Indications of a spatial variation of the fine structure
  constant}},  {\em Phys. Rev. Lett.} {\bf 107} (2011) 191101,
  [\href{http://arxiv.org/abs/1008.3907}{{\tt arXiv:1008.3907}}].

\bibitem{F}
A.~Songaila and L.~L. Cowie, {\it {Constraining the Variation of the Fine
  Structure Constant with Observations of Narrow Quasar Absorption Lines}},
  {\em Astrophys. J.} {\bf 793} (2014) 103,
  [\href{http://arxiv.org/abs/1406.3628}{{\tt arXiv:1406.3628}}].

\bibitem{G}
P.~Molaro, D.~Reimers, I.~I. Agafonova, and S.~A. Levshakov, {\it {Bounds on
  the fine structure constant variability from FeII absorption lines in QSO
  spectra}},  {\em Eur. Phys. J. ST} {\bf 163} (2008) 173--189,
  [\href{http://arxiv.org/abs/0712.4380}{{\tt arXiv:0712.4380}}].

\bibitem{H}
H.~Chand, R.~Srianand, P.~Petitjean, B.~Aracil, R.~Quast, and D.~Reimers, {\it
  {On the variation of the fine-structure constant: Very high resolution
  spectrum of QSO HE 0515-4414}},  {\em Astron. Astrophys.} {\bf 451} (2006)
  45--56, [\href{http://arxiv.org/abs/astro-ph/0601194}{{\tt
  astro-ph/0601194}}].

\bibitem{I}
I.~I. Agafonova, P.~Molaro, S.~A. Levshakov, and J.~L. Hou, {\it {First
  measurement of Mg isotope abundances at high redshifts and accurate estimate
  of $\Delta \alpha/\alpha$}},  {\em Astron. Astrophys.} {\bf 529} (2011) A28,
  [\href{http://arxiv.org/abs/1102.2967}{{\tt arXiv:1102.2967}}].

\bibitem{J}
P.~Molaro et~al., {\it {The UVES Large Program for Testing Fundamental Physics:
  I Bounds on a change in alpha towards quasar HE 2217-2818}},  {\em Astron.
  Astrophys.} {\bf 555} (2013) A68, [\href{http://arxiv.org/abs/1305.1884}{{\tt
  arXiv:1305.1884}}].

\bibitem{K}
T.~M. Evans et~al., {\it {The UVES Large Program for testing fundamental
  physics – III. Constraints on the fine-structure constant from three
  telescopes}},  {\em Mon. Not. Roy. Astron. Soc.} {\bf 445} (2014), no.~1
  128--150, [\href{http://arxiv.org/abs/1409.1923}{{\tt arXiv:1409.1923}}].

\bibitem{N}
M.~T. Murphy, J.~K. Webb, V.~V. Flambaum, M.~J. Drinkwater, F.~Combes, and
  T.~Wiklind, {\it {Improved constraints on possible variation of physical
  constants from H I 21cm and molecular QSO absorption lines}},  {\em Mon. Not.
  Roy. Astron. Soc.} {\bf 327} (2001) 1244,
  [\href{http://arxiv.org/abs/astro-ph/0101519}{{\tt astro-ph/0101519}}].

\bibitem{King}
J.~A. King, J.~K. Webb, M.~T. Murphy, V.~V. Flambaum, R.~F. Carswell, M.~B.
  Bainbridge, M.~R. Wilczynska, and F.~E. Koch, {\it {Spatial variation in the
  fine-structure constant -- new results from VLT/UVES}},  {\em Mon. Not. Roy.
  Astron. Soc.} {\bf 422} (2012) 3370--3413,
  [\href{http://arxiv.org/abs/1202.4758}{{\tt arXiv:1202.4758}}].

\bibitem{Rahmani}
H.~Rahmani, R.~Srianand, N.~Gupta, P.~Petitjean, P.~Noterdaeme, and D.~A.
  V\'{a}squez, {\it {Constraining the variation of fundamental constants at z
  $\sim1.3$ using 21-cm absorbers}},  {\em Mon. Not. Roy. Astron. Soc.} {\bf
  425} (2012) 556--576, [\href{http://arxiv.org/abs/1206.2653}{{\tt
  arXiv:1206.2653}}].

\bibitem{Murphy04}
M.~T. Murphy, V.~V. Flambaum, J.~K. Webb, V.~V. Dzuba, J.~X. Prochaska, and
  A.~M. Wolfe, {\it {Constraining variations in the fine - structure constant,
  quark masses and the strong interaction}},  {\em Lect. Notes Phys.} {\bf 648}
  (2004) 131--150, [\href{http://arxiv.org/abs/astro-ph/0310318}{{\tt
  astro-ph/0310318}}].

\bibitem{AD}
M.~T. Murphy, J.~K. Webb, V.~V. Flambaum, J.~X. Prochaska, and A.~M. Wolfe,
  {\it {Further constraints on variation of the fine structure constant from
  alkali doublet QSO absorption lines}},  {\em Mon. Not. Roy. Astron. Soc.}
  {\bf 327} (2001) 1237, [\href{http://arxiv.org/abs/astro-ph/0012421}{{\tt
  astro-ph/0012421}}].

\bibitem{O}
K.~A. Olive, M.~Pospelov, Y.-Z. Qian, G.~Manhes, E.~Vangioni-Flam, A.~Coc, and
  M.~Casse, {\it {A Re-examination of the Re-187 bound on the variation of
  fundamental couplings}},  {\em Phys. Rev.} {\bf D69} (2004) 027701,
  [\href{http://arxiv.org/abs/astro-ph/0309252}{{\tt astro-ph/0309252}}].

\bibitem{Fujii}
Y.~Fujii, A.~Iwamoto, T.~Fukahori, T.~Ohnuki, M.~Nakagawa, H.~Hidaka, Y.~Oura,
  and P.~Moller, {\it {Nuclear data in Oklo and time variability of fundamental
  coupling constants}},  in {\em {International Conference on Nuclear Data for
  Science and Technology (ND 2001): Embracing the Future at the Beginning of
  the 21st Century Tsukuba, Japan, October 7-12, 2001}}, 2002.
\newblock \href{http://arxiv.org/abs/hep-ph/0205206}{{\tt hep-ph/0205206}}.

\bibitem{Davis}
E.~D. Davis and L.~Hamdan, {\it {Reappraisal of the limit on the variation in
  $\alpha$ implied by the Oklo natural fission reactors}},  {\em Phys. Rev.}
  {\bf C92} (2015), no.~1 014319, [\href{http://arxiv.org/abs/1503.06011}{{\tt
  arXiv:1503.06011}}].

\bibitem{M}
T.~Rosenband, D.~Hume, P.~Schmidt, C.~Chou, A.~Brusch, L.~Lorini, W.~Oskay,
  R.~Drullinger, T.~Fortier, J.~Stalnaker, S.~Diddams, W.~Swann, N.~Newbury,
  W.~Itano, D.~Wineland, and B.~J., {\it {Frequency Ratio of $\text{Al}^{+}$
  and $\text{Hg}^{+}$ Single-Ion Optical Clocks; Metrology at the 17th Decimal
  Place}},  {\em Science} {\bf 319} (2008), no.~5871 1808.

\bibitem{Bekenstein}
J.~D. Bekenstein, {\it {The Relation between physical and gravitational
  geometry}},  {\em Phys. Rev.} {\bf D48} (1993) 3641--3647,
  [\href{http://arxiv.org/abs/gr-qc/9211017}{{\tt gr-qc/9211017}}].

\bibitem{Koivisto}
T.~Koivisto, D.~Wills, and I.~Zavala, {\it {Dark D-brane Cosmology}},  {\em
  JCAP} {\bf 1406} (2014) 036, [\href{http://arxiv.org/abs/1312.2597}{{\tt
  arXiv:1312.2597}}].

\bibitem{Kaloper}
N.~Kaloper, {\it {Disformal inflation}},  {\em Phys. Lett.} {\bf B583} (2004)
  1--13, [\href{http://arxiv.org/abs/hep-ph/0312002}{{\tt hep-ph/0312002}}].

\bibitem{Clayton}
M.~A. Clayton and J.~W. Moffat, {\it {Dynamical mechanism for varying light
  velocity as a solution to cosmological problems}},  {\em Phys. Lett.} {\bf
  B460} (1999) 263--270, [\href{http://arxiv.org/abs/astro-ph/9812481}{{\tt
  astro-ph/9812481}}].

\bibitem{Magueijo}
J.~Magueijo, {\it {New varying speed of light theories}},  {\em Rept. Prog.
  Phys.} {\bf 66} (2003) 2025,
  [\href{http://arxiv.org/abs/astro-ph/0305457}{{\tt astro-ph/0305457}}].

\bibitem{Bassett}
B.~A. Bassett, S.~Liberati, C.~Molina-Paris, and M.~Visser, {\it
  {Geometrodynamics of variable speed of light cosmologies}},  {\em Phys. Rev.}
  {\bf D62} (2000) 103518, [\href{http://arxiv.org/abs/astro-ph/0001441}{{\tt
  astro-ph/0001441}}].

\bibitem{Rham1}
C.~de~Rham and G.~Gabadadze, {\it {Generalization of the Fierz-Pauli Action}},
  {\em Phys. Rev.} {\bf D82} (2010) 044020,
  [\href{http://arxiv.org/abs/1007.0443}{{\tt arXiv:1007.0443}}].

\bibitem{Rham2}
C.~de~Rham, G.~Gabadadze, and A.~J. Tolley, {\it {Resummation of Massive
  Gravity}},  {\em Phys. Rev. Lett.} {\bf 106} (2011) 231101,
  [\href{http://arxiv.org/abs/1011.1232}{{\tt arXiv:1011.1232}}].

\bibitem{Koivisto1}
T.~S. Koivisto, {\it {Disformal quintessence}},
  \href{http://arxiv.org/abs/0811.1957}{{\tt arXiv:0811.1957}}.

\bibitem{Zuma5}
M.~Zumalac\'{a}rregui, T.~S. Koivisto, D.~F. Mota, and P.~Ruiz-Lapuente, {\it
  {Disformal Scalar Fields and the Dark Sector of the Universe}},  {\em JCAP}
  {\bf 1005} (2010) 038, [\href{http://arxiv.org/abs/1004.2684}{{\tt
  arXiv:1004.2684}}].

\bibitem{Sakstein:2014aca}
J.~Sakstein, {\it {Towards Viable Cosmological Models of Disformal Theories of
  Gravity}},  {\em Phys. Rev.} {\bf D91} (2015), no.~2 024036,
  [\href{http://arxiv.org/abs/1409.7296}{{\tt arXiv:1409.7296}}].

\bibitem{Sakstein:2014isa}
J.~Sakstein, {\it {Disformal Theories of Gravity: From the Solar System to
  Cosmology}},  {\em JCAP} {\bf 1412} (2014) 012,
  [\href{http://arxiv.org/abs/1409.1734}{{\tt arXiv:1409.1734}}].

\bibitem{Bettoni:2013diz}
D.~Bettoni and S.~Liberati, {\it {Disformal invariance of second order
  scalar-tensor theories: Framing the Horndeski action}},  {\em Phys. Rev.}
  {\bf D88} (2013) 084020, [\href{http://arxiv.org/abs/1306.6724}{{\tt
  arXiv:1306.6724}}].

\bibitem{Brax:2014vva}
P.~Brax and C.~Burrage, {\it {Constraining Disformally Coupled Scalar Fields}},
   {\em Phys. Rev.} {\bf D90} (2014), no.~10 104009,
  [\href{http://arxiv.org/abs/1407.1861}{{\tt arXiv:1407.1861}}].

\bibitem{MM}
A.~Hees, O.~Minazzoli, and J.~Larena, {\it {Breaking of the equivalence
  principle in the electromagnetic sector and its cosmological signatures}},
  {\em Phys. Rev.} {\bf D90} (2014) 124064,
  [\href{http://arxiv.org/abs/1406.6187}{{\tt arXiv:1406.6187}}].

\bibitem{Bekenstein1}
J.~D. Bekenstein, {\it {Fine Structure Constant: Is It Really a Constant?}},
  {\em Phys. Rev.} {\bf D25} (1982) 1527--1539.

\bibitem{Barrow}
H.~B. Sandvik, J.~D. Barrow, and J.~Magueijo, {\it {A simple cosmology with a
  varying fine structure constant}},  {\em Phys. Rev. Lett.} {\bf 88} (2002)
  031302, [\href{http://arxiv.org/abs/astro-ph/0107512}{{\tt
  astro-ph/0107512}}].

\bibitem{Barrow1}
J.~D. Barrow and S.~Z.~W. Lip, {\it {A Generalized Theory of Varying Alpha}},
  {\em Phys. Rev.} {\bf D85} (2012) 023514,
  [\href{http://arxiv.org/abs/1110.3120}{{\tt arXiv:1110.3120}}].

\bibitem{Barrow2}
J.~D. Barrow and A.~A.~H. Graham, {\it {General Dynamics of Varying-Alpha
  Universes}},  {\em Phys. Rev.} {\bf D88} (2013) 103513,
  [\href{http://arxiv.org/abs/1307.6816}{{\tt arXiv:1307.6816}}].

\bibitem{Minazzoli2}
O.~Minazzoli, {\it {Conservation laws in theories with universal gravity/matter
  coupling}},  {\em Phys. Rev.} {\bf D88} (2013) 027506,
  [\href{http://arxiv.org/abs/1307.1590}{{\tt arXiv:1307.1590}}].

\bibitem{Minazzoli1}
O.~Minazzoli and A.~Hees, {\it {Late-time cosmology of a scalar-tensor theory
  with a universal multiplicative coupling between the scalar field and the
  matter Lagrangian}},  {\em Phys. Rev.} {\bf D90} (2014), no.~2 023017,
  [\href{http://arxiv.org/abs/1404.4266}{{\tt arXiv:1404.4266}}].

\bibitem{Etherington1}
I.~Etherington, {\it {On the definition of distance in general relativity}},
  {\em Philos. Mag.} {\bf 15} (1933), no.~7 761.

\bibitem{Etherington2}
I.~Etherington, {\it {Republication of: LX. On the definition of distance in
  general relativity}},  {\em Gen. Relativ. Gravit.} {\bf 39} (2007), no.~7
  1055.

\bibitem{Ellis1}
G.~Ellis, {\it {On the definition of distance in general relativity: I. M. H.
  Etherington (Philosophical Magazine ser. 7, vol. 15, 761 (1933))}},  {\em
  Gen. Relativ. Gravit.} {\bf 39} (2007), no.~7 1047.

\bibitem{Ellis2}
G.~F.~R. Ellis, {\it {Relativistic cosmology}},  {\em Gen. Rel. Grav.} {\bf 41}
  (2009) 581--660. [Proc. Int. Sch. Phys. Fermi47,104(1971)].

\bibitem{Ellis3}
G.~F.~R. Ellis, R.~Poltis, J.-P. Uzan, and A.~Weltman, {\it {Blackness of the
  cosmic microwave background spectrum as a probe of the distance-duality
  relation}},  {\em Phys. Rev.} {\bf D87} (2013), no.~10 103530,
  [\href{http://arxiv.org/abs/1301.1312}{{\tt arXiv:1301.1312}}].

\bibitem{Lima1}
J.~A.~S. Lima, {\it {Thermodynamics of decaying vacuum cosmologies}},  {\em
  Phys. Rev.} {\bf D54} (1996) 2571--2577,
  [\href{http://arxiv.org/abs/gr-qc/9605055}{{\tt gr-qc/9605055}}].

\bibitem{Lima2}
J.~A.~S. Lima, A.~I. Silva, and S.~M. Viegas, {\it {Is the radiation
  temperature redshift relation of the standard cosmology in accordance with
  the data?}},  {\em Mon. Not. Roy. Astron. Soc.} {\bf 312} (2000) 747--752.

\bibitem{Carsten}
C.~van~de Bruck, J.~Morrice, and S.~Vu, {\it {Constraints on Nonconformal
  Couplings from the Properties of the Cosmic Microwave Background Radiation}},
   {\em Phys. Rev. Lett.} {\bf 111} (2013) 161302,
  [\href{http://arxiv.org/abs/1303.1773}{{\tt arXiv:1303.1773}}].

\bibitem{Brax}
P.~Brax, C.~Burrage, A.-C. Davis, and G.~Gubitosi, {\it {Cosmological Tests of
  the Disformal Coupling to Radiation}},  {\em JCAP} {\bf 1311} (2013) 001,
  [\href{http://arxiv.org/abs/1306.4168}{{\tt arXiv:1306.4168}}].

\bibitem{Dam1}
T.~Damour, F.~Piazza, and G.~Veneziano, {\it {Violations of the equivalence
  principle in a dilaton runaway scenario}},  {\em Phys. Rev.} {\bf D66} (2002)
  046007, [\href{http://arxiv.org/abs/hep-th/0205111}{{\tt hep-th/0205111}}].

\bibitem{Dam2}
T.~Damour, F.~Piazza, and G.~Veneziano, {\it {Runaway dilaton and equivalence
  principle violations}},  {\em Phys. Rev. Lett.} {\bf 89} (2002) 081601,
  [\href{http://arxiv.org/abs/gr-qc/0204094}{{\tt gr-qc/0204094}}].

\bibitem{Dam3}
T.~Damour and A.~M. Polyakov, {\it {The String dilaton and a least coupling
  principle}},  {\em Nucl. Phys.} {\bf B423} (1994) 532--558,
  [\href{http://arxiv.org/abs/hep-th/9401069}{{\tt hep-th/9401069}}].

\bibitem{Dam4}
T.~Damour and A.~M. Polyakov, {\it {String theory and gravity}},  {\em Gen.
  Rel. Grav.} {\bf 26} (1994) 1171--1176,
  [\href{http://arxiv.org/abs/gr-qc/9411069}{{\tt gr-qc/9411069}}].

\bibitem{Nunes1}
E.~J. Copeland, N.~J. Nunes, and M.~Pospelov, {\it {Models of quintessence
  coupled to the electromagnetic field and the cosmological evolution of
  alpha}},  {\em Phys. Rev.} {\bf D69} (2004) 023501,
  [\href{http://arxiv.org/abs/hep-ph/0307299}{{\tt hep-ph/0307299}}].

\bibitem{Olive}
K.~A. Olive and M.~Pospelov, {\it {Evolution of the fine structure constant
  driven by dark matter and the cosmological constant}},  {\em Phys. Rev.} {\bf
  D65} (2002) 085044, [\href{http://arxiv.org/abs/hep-ph/0110377}{{\tt
  hep-ph/0110377}}].

\bibitem{Nunes2}
N.~J. Nunes and J.~E. Lidsey, {\it {Reconstructing the dark energy equation of
  state with varying alpha}},  {\em Phys. Rev.} {\bf D69} (2004) 123511,
  [\href{http://arxiv.org/abs/astro-ph/0310882}{{\tt astro-ph/0310882}}].

\bibitem{Uzan}
J.-P. Uzan, {\it {Varying Constants, Gravitation and Cosmology}},  {\em Living
  Rev. Rel.} {\bf 14} (2011) 2, [\href{http://arxiv.org/abs/1009.5514}{{\tt
  arXiv:1009.5514}}].

\bibitem{Zuma1}
M.~Zumalac\'{a}rregui and J.~Garc\'{i}a-Bellido, {\it {Transforming gravity:
  from derivative couplings to matter to second-order scalar-tensor theories
  beyond the Horndeski Lagrangian}},  {\em Phys. Rev.} {\bf D89} (2014) 064046,
  [\href{http://arxiv.org/abs/1308.4685}{{\tt arXiv:1308.4685}}].

\bibitem{Zuma2}
M.~Zumalac\'{a}rregui, T.~S. Koivisto, and D.~F. Mota, {\it {DBI Galileons in
  the Einstein Frame: Local Gravity and Cosmology}},  {\em Phys. Rev.} {\bf
  D87} (2013) 083010, [\href{http://arxiv.org/abs/1210.8016}{{\tt
  arXiv:1210.8016}}].

\bibitem{Minamitsuji}
M.~Minamitsuji, {\it {Disformal transformation of cosmological perturbations}},
   {\em Phys. Lett.} {\bf B737} (2014) 139--150,
  [\href{http://arxiv.org/abs/1409.1566}{{\tt arXiv:1409.1566}}].

\bibitem{Zuma3}
T.~S. Koivisto, D.~F. Mota, and M.~Zumalac\'{a}rregui, {\it {Screening
  Modifications of Gravity through Disformally Coupled Fields}},  {\em Phys.
  Rev. Lett.} {\bf 109} (2012) 241102,
  [\href{http://arxiv.org/abs/1205.3167}{{\tt arXiv:1205.3167}}].

\bibitem{Jack}
C.~van~de Bruck and J.~Morrice, {\it {Disformal couplings and the dark sector
  of the universe}},  {\em JCAP} {\bf 1504} (2015), no.~04 036,
  [\href{http://arxiv.org/abs/1501.03073}{{\tt arXiv:1501.03073}}].

\bibitem{Suzuki}
N.~Suzuki et~al., {\it {The Hubble Space Telescope Cluster Supernova Survey: V.
  Improving the Dark Energy Constraints Above $z>1$ and Building an
  Early-Type-Hosted Supernova Sample}},  {\em Astrophys. J.} {\bf 746} (2012)
  85, [\href{http://arxiv.org/abs/1105.3470}{{\tt arXiv:1105.3470}}].

\bibitem{Khoury}
S.~Das, P.~S. Corasaniti, and J.~Khoury, {\it {Super-acceleration as signature
  of dark sector interaction}},  {\em Phys. Rev.} {\bf D73} (2006) 083509,
  [\href{http://arxiv.org/abs/astro-ph/0510628}{{\tt astro-ph/0510628}}].

\bibitem{P2}
{\bf Planck} Collaboration, P.~A.~R. Ade et~al., {\it {Planck 2015 results.
  XIII. Cosmological parameters}},  \href{http://arxiv.org/abs/1502.01589}{{\tt
  arXiv:1502.01589}}.

\bibitem{Chiba}
T.~Chiba, {\it {The Constancy of the Constants of Nature: Updates}},  {\em
  Prog. Theor. Phys.} {\bf 126} (2011) 993--1019,
  [\href{http://arxiv.org/abs/1111.0092}{{\tt arXiv:1111.0092}}].

\bibitem{Dvali}
G.~R. Dvali and M.~Zaldarriaga, {\it {Changing $\alpha$ with time: Implications
  for fifth force type experiments and quintessence}},  {\em Phys. Rev. Lett.}
  {\bf 88} (2002) 091303, [\href{http://arxiv.org/abs/hep-ph/0108217}{{\tt
  hep-ph/0108217}}].

\bibitem{Calabrese}
E.~Calabrese, E.~Menegoni, C.~J. A.~P. Martins, A.~Melchiorri, and G.~Rocha,
  {\it {Constraining Variations in the Fine Structure Constant in the presence
  of Early Dark Energy}},  {\em Phys. Rev.} {\bf D84} (2011) 023518,
  [\href{http://arxiv.org/abs/1104.0760}{{\tt arXiv:1104.0760}}].

\bibitem{Pinho}
C.~J. A.~P. Martins and A.~M.~M. Pinho, {\it {Fine-structure constant
  constraints on dark energy}},  {\em Phys. Rev.} {\bf D91} (2015), no.~10
  103501, [\href{http://arxiv.org/abs/1505.02196}{{\tt arXiv:1505.02196}}].

\bibitem{ESPRESSO}
F.~Pepe et~al., {\it {ESPRESSO — An Echelle SPectrograph for Rocky
  Exo-planets Search and Stable Spectroscopic Observations}},  {\em The
  Messenger (ESO)} {\bf 153} (2013) 6.

\bibitem{ELT1}
ESO, ``The e-elt construction proposal.''
  \url{http://www.eso.org/public/products/books/book_0046}, 2011.
\newblock Accessed: 25/09/2015.

\bibitem{ELT2}
R.~Maiolino et~al., {\it {A Community Science Case for E-ELT HIRES}},
  \href{http://arxiv.org/abs/1310.3163}{{\tt arXiv:1310.3163}}.

\bibitem{Sandage1}
A.~Sandage, {\it {The Change of Redshift and Apparent Luminosity of Galaxies
  due to the Deceleration of Selected Expanding Universes.}},  {\em Astrophys.
  J.} {\bf 136} (1962) 319.

\bibitem{Sandage2}
A.~Loeb, {\it {Direct Measurement of Cosmological Parameters from the Cosmic
  Deceleration of Extragalactic Objects}},  {\em Astrophys. J.} {\bf 499}
  (1998) L111--L114, [\href{http://arxiv.org/abs/astro-ph/9802122}{{\tt
  astro-ph/9802122}}].

\bibitem{Sandage3}
J.~Liske et~al., {\it {Cosmic dynamics in the era of Extremely Large
  Telescopes}},  {\em Mon. Not. Roy. Astron. Soc.} {\bf 386} (2008) 1192--1218,
  [\href{http://arxiv.org/abs/0802.1532}{{\tt arXiv:0802.1532}}].

\bibitem{Sandage4}
P.~E. Vielzeuf and C.~J. A.~P. Martins, {\it {Probing dark energy beyond $z=2$
  with CODEX}},  {\em Phys. Rev.} {\bf D85} (2012) 087301,
  [\href{http://arxiv.org/abs/1202.4364}{{\tt arXiv:1202.4364}}].

\bibitem{ALMA1}
V.~Fish et~al., {\it {High-Angular-Resolution and High-Sensitivity Science
  Enabled by Beamformed ALMA}},  \href{http://arxiv.org/abs/1309.3519}{{\tt
  arXiv:1309.3519}}.

\bibitem{ALMA2}
R.~P.~J. Tilanus et~al., {\it {Future mmVLBI Research with ALMA: A European
  vision}},  \href{http://arxiv.org/abs/1406.4650}{{\tt arXiv:1406.4650}}.

\bibitem{Martins}
C.~J. A.~P. Martins, {\it {Fundamental cosmology in the E-ELT era: The status
  and future role of tests of fundamental coupling stability}},  {\em Gen. Rel.
  Grav.} {\bf 47} (2015), no.~1 1843,
  [\href{http://arxiv.org/abs/1412.0108}{{\tt arXiv:1412.0108}}].

\bibitem{white_dwarf1}
J.~Bagdonaite, E.~J. Salumbides, S.~P. Preval, M.~A. Barstow, J.~D. Barrow,
  M.~T. Murphy, and W.~Ubachs, {\it {Limits on a Gravitational Field Dependence
  of the Proton-Electron Mass Ratio from H$_2$ in White Dwarf Stars}},  {\em
  Phys. Rev. Lett.} {\bf 113} (2014), no.~12 123002,
  [\href{http://arxiv.org/abs/1409.1000}{{\tt arXiv:1409.1000}}].

\bibitem{white_dwarf2}
J.~C. Berengut, V.~V. Flambaum, A.~Ong, J.~K. Webb, J.~D. Barrow, M.~A.
  Barstow, S.~P. Preval, and J.~B. Holberg, {\it {Limits on the dependence of
  the fine-structure constant on gravitational potential from white-dwarf
  spectra}},  {\em Phys. Rev. Lett.} {\bf 111} (2013), no.~1 010801,
  [\href{http://arxiv.org/abs/1305.1337}{{\tt arXiv:1305.1337}}].

\end{thebibliography}\endgroup

\end{document}